\documentclass{emulateapj}
\usepackage{natbib,aas_macros}
\usepackage{amsmath,graphicx,psfrag,float,color,booktabs}
\usepackage{ulem}

\providecommand{\inst}{\altaffilmark}

\newcommand{\tyc}{TYC 3010}
\newcommand{\kms}{km~s$^{-1}$}

\begin{document}

\submitted{Published in The Astronomical Journal, 145:139, 2013 May}

\title{A cautionary tale: MARVELS brown dwarf candidate reveals itself to be a 
very long period, highly eccentric spectroscopic stellar binary}

%%%%%%%%%%%%%%%%%%%%%%%%%%%%%%%%
%%% AUTHORS %%%%
%%%%%%%%%%%%%%%%%%%%%%%%%%%%%%%%
\author{
Claude~E.~Mack~III\inst{1},
Jian~Ge\inst{2},
Rohit~Deshpande\inst{3,4},
John~P.~Wisniewski\inst{5},
Keivan~G.~Stassun\inst{1,6},
B.~Scott~Gaudi\inst{7},
Scott~W.~Fleming\inst{2,3,4},
Suvrath~Mahadevan\inst{3,4},
Nathan~De~Lee\inst{2,1},
Jason~Eastman\inst{7},
Luan~Ghezzi\inst{8,9},
Jonay~I.~Gonz\'{a}lez~Hern\'{a}ndez\inst{10,11},
Bruno~Femen\'{i}a\inst{10,11},
Let\'{i}cia~Ferreira\inst{12,9},
Gustavo~Porto~de~Mello\inst{12,9},
Justin~R.~Crepp\inst{13},
Daniel~Mata~S\'{a}nchez\inst{10,11},
Eric~Agol\inst{14},
Thomas~G.~Beatty\inst{7},
Dmitry~Bizyaev\inst{15},
Howard~Brewington\inst{15},
Phillip~A.~Cargile\inst{1},
Luiz N.~da~Costa\inst{8,9},
Massimiliano~Esposito\inst{10,11},
Garret~Ebelke\inst{15},
Leslie~Hebb\inst{1,14},
Peng~Jiang\inst{2},
Stephen~R.~Kane\inst{16},
Brian~Lee\inst{2},
Marcio~A.~G.~Maia\inst{8,9},
Elena~Malanushenko\inst{15},
Victor~Malanushenko\inst{15},
Daniel~Oravetz\inst{15},
Martin~Paegert\inst{1},
Kaike~Pan\inst{15},
Carlos~Allende~Prieto\inst{10,11},
Joshua~Pepper\inst{1},
Rafael~Rebolo\inst{10,11},
Arpita~Roy\inst{3},
Bas\'{i}lio~X.~Santiago\inst{17,9},
Donald~P.~Schneider\inst{3,4},
Audrey~Simmons\inst{15},
Robert~J.~Siverd\inst{1},
Stephanie~Snedden\inst{15},
and Benjamin~M.~Tofflemire\inst{18}
}  

%%%%%%%%%%%%%%%%%%%%%%%%%%%%%%%%
%%% AFFILIATIONS %%%%
%%%%%%%%%%%%%%%%%%%%%%%%%%%%%%%%
\affil{
\mbox{$^{1}$ {Department of Physics and Astronomy, Vanderbilt University, Nashville, TN 37235, USA; claude.e.mack@vanderbilt.edu}}\\
\mbox{$^{2}$ {Department of Astronomy, University of Florida, 211 Bryant Space Science Center, Gainesville, FL, 32611-2055, USA}}\\
\mbox{$^{3}$ {Department of Astronomy and Astrophysics, The Pennsylvania State University, University Park, PA 16802, USA}}\\
\mbox{$^{4}$ {Center for Exoplanets and Habitable Worlds, Pennsylvania State University, University Park, PA 16802, USA}}\\
\mbox{$^{5}$ {Homer L Dodge Department of Physics \& Astronomy, University of Oklahoma, 440 W Brooks St, Norman, OK 73019, USA}}\\
\mbox{$^{6}$ {Department of Physics, Fisk University, Nashville, TN, USA}}\\
\mbox{$^{7}$ {Department of Astronomy, The Ohio State University, 140 West 18th Avenue, Columbus, OH 43210, USA}}\\
\mbox{$^{8}$ {Observat\'{o}rio Nacional, Rua Gal. Jos\'{e} Cristino 77, Rio de Janeiro, RJ 20921-400, Brazil}}\\
\mbox{$^{9}$ {Laborat\'{o}rio Interinstitucional de e-Astronomia--LIneA, Rio de Janeiro, RJ 20921-400, Brazil}}\\
\mbox{$^{10}$ {Instituto de Astrof\'{i}sica de Canarias (IAC), E-38205 La Laguna, Tenerife, Spain}}\\
\mbox{$^{11}$ {Departamento de Astrof\'{i}sica, Universidad de La Laguna, E-38206 La Laguna, Tenerife, Spain}}\\
\mbox{$^{12}$ {Observat\'{o}rio do Valongo, Universidade Federal do Rio de Janeiro, Rio de Janeiro, RJ 20080-090, Brazil}}\\
\mbox{$^{13}$ {Department of Physics, University of Notre Dame, 225 Nieuwland Science Hall, Notre Dame, IN 46556, USA}}\\
\mbox{$^{14}$ {Astronomy Department, University of Washington, Box 351580, Seattle, WA 98195, USA}}\\
\mbox{$^{15}$ {Apache Point Observatory, P. O. Box 59, Sunspot, NM 88349-0059, USA}}\\
\mbox{$^{16}$ {NASA Exoplanet Science Institute, Caltech, MS 100-22, 770 South Wilson Avenue, Pasadena, CA 91125, USA}}\\
\mbox{$^{17}$ {Instituto de F\'{i}sica, UFRGS, Caixa Postal 15051, Porto Alegre, RS 91501-970, Brazil}}\\
\mbox{$^{18}$ {Astronomy Department, University of Wisconsin-Madison, 475 N Charter St, Madison, WI 53706, USA}}
}

%%%%%%%%%%%%%%%%%%%%%%%%%%%%%%%%
%% ABSTRACT
%%%%%%%%%%%%%%%%%%%%%%%%%%%%%%%%
\begin{abstract}
We report the discovery of a highly eccentric,
double-lined spectroscopic binary star system (TYC~3010-1494-1), comprising two solar-type stars 
that we had initially identified as a single star with a brown dwarf companion. 
At the moderate resolving power of 
the MARVELS spectrograph and the spectrographs used for subsequent radial-velocity (RV) 
measurements ($R\lesssim 30,000$), this particular
stellar binary mimics a single-lined binary with an RV signal that would be
induced by a brown dwarf companion ($M\sin{i}\sim50\, {M}_{\rm{Jup}}$)
to a solar-type primary. At least three properties of this  system 
allow it to masquerade as a single star with a very low-mass companion:
its large eccentricity ($e\sim0.8$), its relatively long period ($P\sim238$ days), 
and the approximately perpendicular orientation of the semi-major axis with respect to the 
line of sight ($\omega\sim189^{\circ}$). As a result of these properties, 
for $\sim$95\% of the orbit the two sets of stellar spectral lines are
completely blended, and the RV measurements based on centroiding on the apparently single-lined spectrum is
very well fit by an orbit solution indicative of a brown dwarf companion on a more circular orbit ($e\sim0.3$).
Only during the $\sim$5\% of the orbit near periastron passage does the true, 
double-lined nature and large RV amplitude of $\sim$15 km~s$^{-1}$ reveal itself.
The discovery of this binary system is an important lesson for RV surveys searching
for substellar companions; at a given resolution and observing cadence, a survey will be susceptible to
these kinds of astrophysical false positives for a range of orbital parameters. 
Finally, for surveys like MARVELS that lack the resolution for a useful line
bisector analysis, it is imperative to monitor the peak of the cross-correlation function
for suspicious changes in width or shape, so that such false positives
can be flagged during the candidate vetting process.
\end{abstract}


%%%%%%%%%%%%%%%%%%%%%%%%%%%%%%%%
%% INTRODUCTION
%%%%%%%%%%%%%%%%%%%%%%%%%%%%%%%%
\section{Introduction}
\label{sec:intro}

%%%%%%%%%%%%%%%%%%%%%%%%%
%% FIGURE 1. BD RV CURVE
%%%%%%%%%%%%%%%%%%%%%%%%%
\begin{figure*}
\centering
  \includegraphics[width=4.3in]{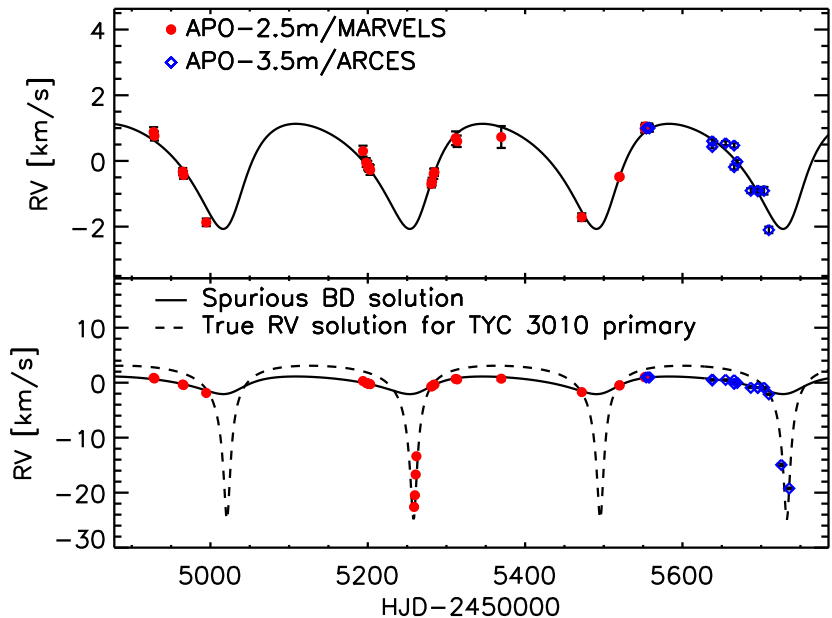}
  \caption{
     The radial velocity data obtained with the MARVELS (red)
     and ARCES (blue) spectrographs at the time that we began to suspect that
     \tyc\ was a double-lined spectroscopic binary (SB2) instead of a brown dwarf (BD) companion
     to a solar-type star. In the top panel, we show the {\sc exofast} fit~\citep[solid line;][]{eg13} to the
     low-amplitude RV variations that are observed when the binary is away from periastron. This solution
     corresponds to a substellar companion in the BD regime ($M\sin{i}\sim 50\, M_{\rm Jup}$) orbiting
     a solar-type primary with a period of $\sim$238 days.
     In the bottom panel, we include the high-amplitude MARVELS (red points near HJD 2455250) 
     and ARCES (blue points near HJD 2455730) outliers that were initially
     thought to be spurious, as well as the final, true RV curve (dashed line) for the primary component of the SB2.
     For both spectrographs, the majority of the data agrees well with the BD solution, and it is tempting to
     suspect the outliers as spurious. However, upon investigating the cross-correlation function (CCF) for these
     outliers, the CCFs show strong evidence for a secondary stellar component (see Figure~\ref{fig:ccfs}).
     With the HET/HRS spectrograph we were able to completely cover periastron and confirm that the system is indeed
     are offset from the dashed curve because these points actually correspond to the flux-weighted 
     average of the true primary and secondary RVs. To perform the double-lined fit for these 
     (apparently) single-lined epochs, we first disentangled the primary and secondary components
     as described in Section~\ref{sec:rv_fitting}.\\[5pt]
  }
  \label{fig:mc10rv_curve_bd}
  \end{figure*}
%%%%%%%%%%%%%%%%%%%%%%%%%

As a part of the third phase of the Sloan Digital Sky Survey~\citep[SDSS-III;][]{ew11},
the MARVELS ({\textit{M}ulti-object \textit{A}PO \textit{R}adial 
\textit{V}elocity \textit{E}xoplanet \textit{L}arge-area \textit{S}urvey})
project is searching for substellar companions by monitoring the 
radial velocities (RVs) of 3330 FGK stars~\citep{gm08,gl09,ge09}. 
This sample size is large enough for the project to
find relatively rare objects, such as brown dwarf (BD) companions to solar-type stars.
The paucity of observed BD companions to solar-type stars with separations of $\lesssim$5 AU
is typically referred to as the BD desert~\citep{mb00}. 
Since the size of the MARVELS sample
allows us to begin to quantify how arid the BD desert may be, any MARVELS discovery
of a BD in the desert (or lack thereof) is a step toward increasing our understanding
of BD formation.

In addition to its large homogeneous target sample, MARVELS differs from other
surveys for substellar companions in two key ways. First, the project employs
a dispersed fixed-delay interferometer~\citep[DFDI;][]{gj02,ge02,ed03,gv06,vg10,wg11}.
Second, it uses a multi-object
spectrograph to observe 60 stars simultaneously~\citep{gl09}. 
The DFDI prototype instrument was used to discover the first extrasolar
planet around HD 102195 in 2006 with this new RV method~\citep{gv06}. The MARVELS DFDI technique combines
an interferometer with a medium resolution spectrograph ($R\sim$12,000)
in order to obtain a precision of $\sim$100 m~s$^{-1}$.
Given its RV precision and survey design to monitor each target
with at least 24 RV measurements over at least 1 yr, MARVELS is sensitive 
to BD and low-mass stellar companions with periods ranging
from a few days to hundreds of days. Nonetheless, certain specific types of astrophysical false
positives can mimic substellar companions unless additional vetting is performed. 
This paper describes just such a case, TYC-3010-1494-1 (hereafter \tyc), a stellar binary that 
initially appeared as a single star with a substellar companion and that, through a confluence of 
orbital parameters, continued to masquerade as such 
despite a disconcertingly extensive amount of
observation and analysis.

When we began analysis of \tyc, MARVELS and its pilot project had already detected two BD candidates
orbiting late F stars in the BD 
desert~(\citealt[][at present, we have three more candidates in the desert: ]{fg10,lg11}\citealt{ma13,jp13,dl13}).
The MARVELS discovery data indicated that \tyc\ possessed a substellar companion
with a minimum mass of $\sim50\, M_{\rm{Jup}}$ and that it was
on a $\sim$238-day moderately eccentric orbit 
with an RV amplitude of $\sim$1.5~km~s$^{-1}$ (see the top panel of Figure~\ref{fig:mc10rv_curve_bd}). However, 
given the cadence of MARVELS and the period of the orbit, there were significant gaps in the phase coverage
and additional observations with a different spectrograph were required to constrain the RV solution.
Initially, the follow-up data remained fully consistent with the BD companion scenario.
However, during the course of the program, we found two RV points that were shifted by $\sim$20~\kms\
with respect to most of our data; while investigating the source of these anomalous points, 
we realized that a few similar points had been rejected from our MARVELS discovery data 
by the team's outlier rejection procedures (see bottom panel of Figure~\ref{fig:mc10rv_curve_bd}).  
Examining the cross-correlation function (CCF) of the anomalous RV points (in
both the discovery and subsequent data) revealed evidence that there were two components in the CCF, which
suggested that the companion to the primary was most likely a stellar-mass secondary.
Finally, including the initially flagged outlier measurements and disentangling the 
RV measurements of the two components, the system was found to be a nearly equal-mass stellar binary ($q\sim0.88$) on a 
highly eccentric orbit ($e\sim 0.8$). Evidently, for a system like \tyc, it is possible to clip just a few
measurements and obtain an apparently reasonable solution that is convincing
but completely incorrect.

As large scale RV and transit surveys for exoplanets become more common, it is increasingly
inevitable that any and all forms of astrophysical false positives, despite their rarity,
will be found. Indeed, the first BD candidate discovered by the MARVELS project, \mbox{MARVELS-1~\citep{lg11}}, 
appeared to exhibit evidence for an additional planet-mass companion, but turned out instead to likely
be a quadruple system, comprising four stars 
with no detected BD or planetary-mass companion~\citep{wj13}. 
Akin to \tyc, \mbox{MARVELS-1} is a double-lined spectroscopic binary;
the stars have relative RVs which are sufficiently low that
they are always blended, even at the resolution of the Hobby-Eberly Telescope (HET; $R\sim 60,000$ mode).
Thus, with both \mbox{MARVELS-1} and \tyc, we actually measure a flux-weighted mean of
two sets of stellar spectral lines. This flux-weighted mean exhibits
a suppressed velocity shift that mimics a single-lined binary with
a BD secondary. Both systems possess geometries that allow
them to masquerade as less massive systems: \mbox{MARVELS-1} is nearly face-on, which leads to
low projected velocities, while \tyc\ is on a highly elliptical orbit with
a semi-major axis oriented nearly perpendicular to our line of sight.

Similarly, \citet{mt05} describe what at first appeared to be a 
transiting BD companion to an F star from the TRES transit survey, but turned out instead to
be an F star blended with a G+M stellar eclipsing binary. The system that we describe here follows
these unfortunate examples, and is similarly pernicious.

In the following sections, we present our analysis as a kind of 
cautionary tale for other RV surveys to avoid similar false positives.
In Section~\ref{sec:data}, we describe the spectroscopic and photometric data obtained for \tyc.
In Section~\ref{sec:results}, we discuss in detail the nature of the evidence that 
led us to conclude that \tyc\ was an eccentric 
stellar binary instead of a BD companion to a solar-type star. 
We also present the properties we derived for both components of the spectroscopic binary.
In Section~\ref{sec:discussion}, we discuss the circumstances 
that allowed this false positive to masquerade for so long and
through several vetting steps as a compelling detection of a substellar companion,
and we describe methods that the MARVELS team and other RV surveys can use to recognize this kind of
astrophysical false positive in the future.  
Finally, in Section~\ref{sec:summary}, we conclude with a summary of the main results.
%%%%%%%%%%%%%%%%%%%%%%%%%%%%%%%%
%% END OF INTRODUCTION
%%%%%%%%%%%%%%%%%%%%%%%%%%%%%%%%

%%%%%%%%%%%%%%%%%%%%%%%%%%%%%%%%
%% DATA
%%%%%%%%%%%%%%%%%%%%%%%%%%%%%%%%
\section{Observations and Data Processing}
\label{sec:data}
We obtained a total of 65 RV measurements from the Sloan 2.5m, the
APO 3.5m, and the HET 9.2m telescopes. We will briefly summarize the characteristics
of the data from all three telescopes. For more details of
the analysis, please see~\citet{fg10},~\citet{lg11}, and~\citet{wj12}.

\subsection{SDSS-III MARVELS Discovery RV Data}
A total of 28 spectra (see Table~\ref{tbl:sing_rvs})
of \tyc\ were obtained with the Sloan 2.5m telescope~\citep{gj06}
at Apache Point Observatory (APO). 
The multi-fiber MARVELS spectrograph~\citep{gl09} can simultaneously
measure the RVs of 60 stars during each telescope pointing.
Both beams of the interferometer are imaged onto the detector, so each 50-minute
observation results in two fringed spectra in the wavelength range of
$\sim$500--570 nm with a resolving power of $R\sim12,000$.
The MARVELS interferometer delay calibrations are described in~\citet{wg12a,wg12b}.
For more details on how the data were reduced and analyzed to yield RVs, see~\citet{lg11}.

%%%%%%%%%%%%%%%%%%%%%%%%%%%%%%%%
% TABLE 1: SINGLE-LINED RVS
%%%%%%%%%%%%%%%%%%%%%%%%%%%%%%%%
  \begin{deluxetable}{cccc}
  \tabletypesize{\footnotesize}
  \tablecaption{Observed heliocentric single-lined\protect\\ radial velocities for TYC 3010\label{tbl:sing_rvs}}
  \tablewidth{0.49\textwidth}
  \tablehead{\colhead{HJD} & \colhead{Instrument\tablenotemark{a}} & \colhead{RV (\kms)} & \colhead{$\sigma_{\rm{RV}}$ (\kms)}}
  \startdata
  2454927.82470 & M & 62.681 & 0.148\\
  2454928.85061 & M & 62.564 & 0.139\\
  2454964.76792 & M & 61.479 & 0.108\\
  2454965.77714 & M & 61.374 & 0.113\\
  2454994.69536 & M & 59.933 & 0.115\\
  2455193.91250 & M & 62.102 & 0.165\\
  2455197.96727 & M & 61.753 & 0.134\\
  2455198.94828 & M & 61.714 & 0.095\\
  2455199.96552 & M & 61.664 & 0.139\\
  2455200.98947 & M & 61.585 & 0.097\\
  2455201.97760 & M & 61.587 & 0.116\\
  2455202.99063 & M & 61.528 & 0.149\\
  2455258.88272 & M & 39.192 & 0.091\\
  2455259.83118 & M & 41.327 & 0.092\\
  2455260.82412 & M & 45.097 & 0.145\\
  2455261.82050 & M & 48.416 & 0.096\\
  2455280.77587 & M & 61.103 & 0.105\\
  2455280.76844 & M & 61.174 & 0.117\\
  2455283.81484 & M & 61.411 & 0.154\\
  2455284.75054 & M & 61.461 & 0.112\\
  2455311.68421 & M & 62.493 & 0.209\\
  2455313.62591 & M & 62.402 & 0.174\\
  2455369.64423 & M & 62.531 & 0.333\\
  2455551.99403 & M & 62.788 & 0.120\\
  2455552.98222 & M & 62.856 & 0.104\\
  2455553.98561 & M & 62.795 & 0.121\\
  2455556.97163 & M & 62.821 & 0.123\\
  2455557.97465 & M & 62.801 & 0.104\\
  2455471.98302 & A & 60.138 & 0.116\\
  2455519.95995 & A & 61.359 & 0.052\\
  2455519.98157 & A & 61.371 & 0.051\\
  2455637.88366 & A & 62.452 & 0.055\\
  2455637.92209 & A & 62.278 & 0.048\\
  2455654.83350 & A & 62.390 & 0.059\\
  2455665.65219 & A & 62.323 & 0.065\\
  2455665.69165 & A & 61.664 & 0.075\\
  2455669.60113 & A & 61.827 & 0.052\\
  2455686.82409 & A & 60.946 & 0.076\\
  2455695.66512 & A & 60.949 & 0.039\\
  2455695.70529 & A & 60.931 & 0.053\\
  2455703.61994 & A & 60.942 & 0.116\\
  2455709.77767 & A & 59.749 & 0.098\\
  2455903.90846 & H & 62.448 & 0.051\\
  2455917.87269 & H & 62.237 & 0.060\\
  2455928.84083 & H & 61.759 & 0.046\\
  2455940.80855 & H & 61.122 & 0.058\\
  2455946.80490 & H & 60.285 & 0.055\\
  2455950.80134 & H & 59.539 & 0.045\\
  2455953.82447 & A & 58.385 & 0.049\\
  2455954.00566 & H & 58.467 & 0.050\\
  \enddata
  \tablenotetext{a}{Instruments: MARVELS (M), ARCES (A), and HRS (H) spectrographs.}
  \tablecomments{
  The ARCES and HRS RV values
  were measured as absolute heliocentric RVs, while the MARVELS
  discovery data were measured on a relative instrumental scale; the MARVELS RVs have been offset to the 
  same (heliocentric) scale as the ARCES and HRS measurements.}
  \end{deluxetable}
%%%%%%%%%%%%%%%%%%%%%%%%%%%%%%%%

%%%%%%%%%%%%%%%%%%%%%%%%%%%%%%%%
% TABLE 2: DOUBLE-LINED RVS
%%%%%%%%%%%%%%%%%%%%%%%%%%%%%%%%
  \begin{deluxetable*}{cccccc}
  \tabletypesize{\normalsize}
  \tablecolumns{6} 
  \tablecaption{Observed heliocentric double-lined radial velocities for TYC 3010\label{tbl:doub_rvs}}
  \tablehead{\colhead{HJD} & \colhead{Instrument\tablenotemark{a}} & \colhead{RV$_{\rm primary}$ (\kms)}
        & \colhead{$\sigma_{\rm{RV_{\rm primary}}}$ (\kms)} & \colhead{RV$_{\rm secondary}$ (\kms)} & \colhead{$\sigma_{\rm{RV_{\rm secondary}}}$ (\kms)}}
  \startdata
  2455725.68377 & A & 46.012  & 0.167 & 75.222 & 0.257\\
  2455735.62781 & A & 43.197  & 0.251 & 81.056 & 0.175\\
  2455956.76037 & H & 53.788  & 0.030 & 69.104 & 0.063\\
  2455959.78075 & H & 51.409  & 0.025 & 71.807 & 0.055\\
  2455964.75592 & H & 44.163  & 0.026 & 80.066 & 0.055\\
  2455964.83117 & A & 44.884  & 0.071 & 79.700 & 0.372\\
  2455967.75334 & H & 36.755  & 0.030 & 88.389 & 0.062\\
  2455967.82824 & A & 37.684  & 0.076 & 88.651 & 0.471\\
  2455968.74640 & H & 34.456  & 0.025 & 90.966 & 0.054\\
  2455971.73989 & H & 36.359  & 0.029 & 88.685 & 0.060\\
  2455972.97350 & H & 40.368  & 0.025 & 84.354 & 0.054\\
  2455976.73787 & H & 50.072  & 0.024 & 73.253 & 0.051\\
  2455977.71541 & H & 51.788  & 0.028 & 71.580 & 0.058\\
  2455978.71767 & H & 52.990  & 0.026 & 69.840 & 0.056\\
  2455979.71599 & H & 54.160  & 0.029 & 68.450 & 0.062\\
  \enddata
  \tablenotetext{a}{{\footnotesize Instruments: ARCES (A) and HRS (H) spectrographs.}}
  \end{deluxetable*}
  %\vspace{-0.1in}
  %\tablecomments{blah blah}
%\vspace{15pt}
%%%%%%%%%%%%%%%%%%%%%%%%%%%%%%%%

As described below, it proved essential to examine the CCFs
of the individual spectra. However,
performing a cross-correlation on a DFDI spectrum requires a few steps
beyond what one performs for a typical slit or cross-dispersed echelle spectrograph.
In both cases the images are reduced using standard techniques
(bias subtraction, trace correction, flat fielding etc.)
Once a fully processed two-dimensional spectrum has been extracted, there is a divergence
in the techniques. In the case of a normal spectrum, one merely sums the flux in
the slit (channel) direction to produce a one-dimensional spectrum.  This approach is not
possible in the DFDI technique because the fringing pattern will introduce false
fluctuations in total flux if one just sums in the slit direction.  These fluctuations
will be a function of the phase of the fringe pattern in each pixel channel.  To correct
for this effect, a sinusoidal function of the form $A\sin{(wx + b)}+c$ is fit to each pixel column.
For the purposes of cross-correlation the only term of interest is $c$, or the mean flux in each channel.
A one dimensional spectrum is then constructed using the $c$ term in each channel.
From this point forward the CCF is determined using standard techniques.

\subsection{APO-3.5m/ARCES RV Data}
\label{sec:arces_data}
\begin{figure}[H]
\begin{center}
  \includegraphics[width=0.33\textwidth]{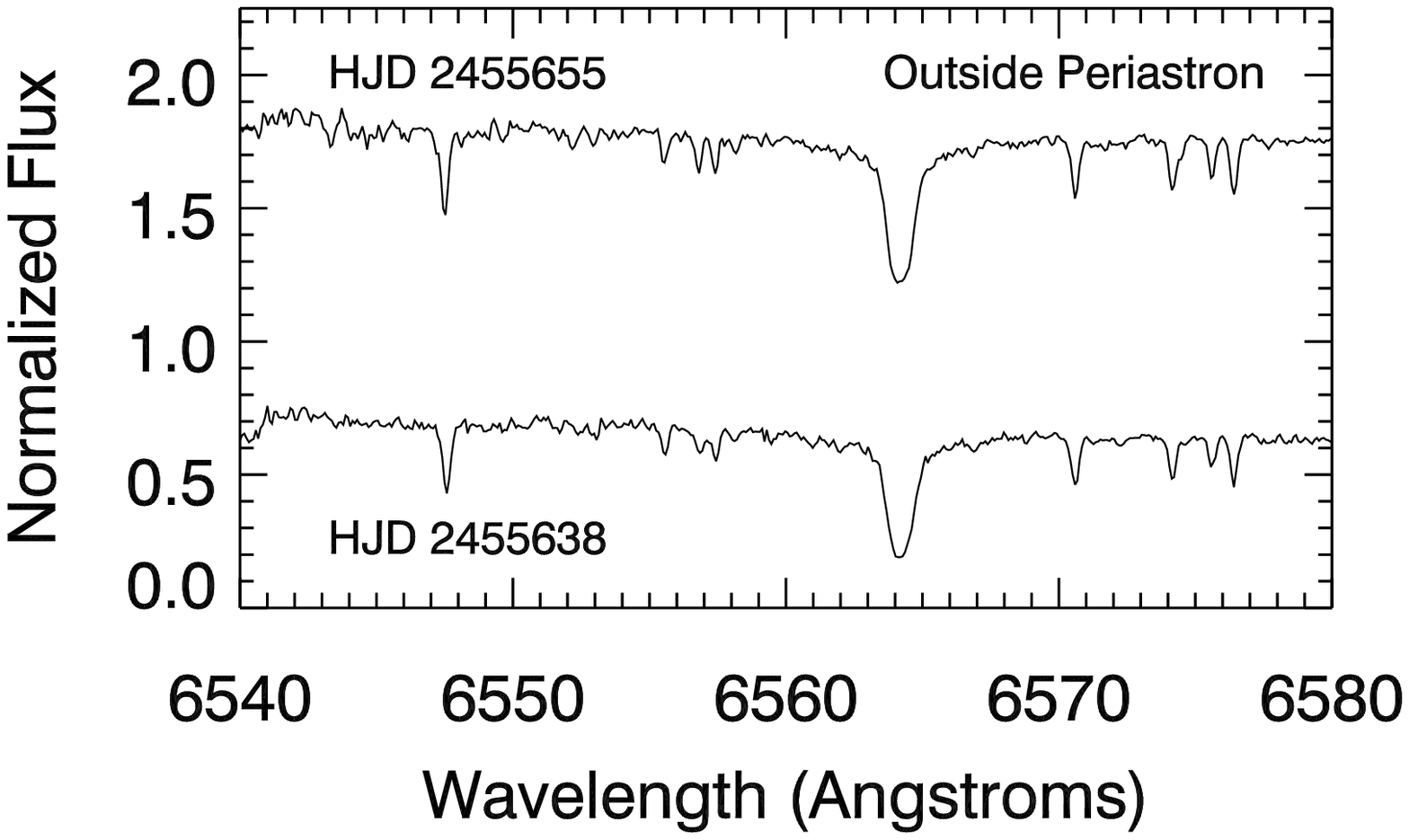}
  \includegraphics[width=0.33\textwidth]{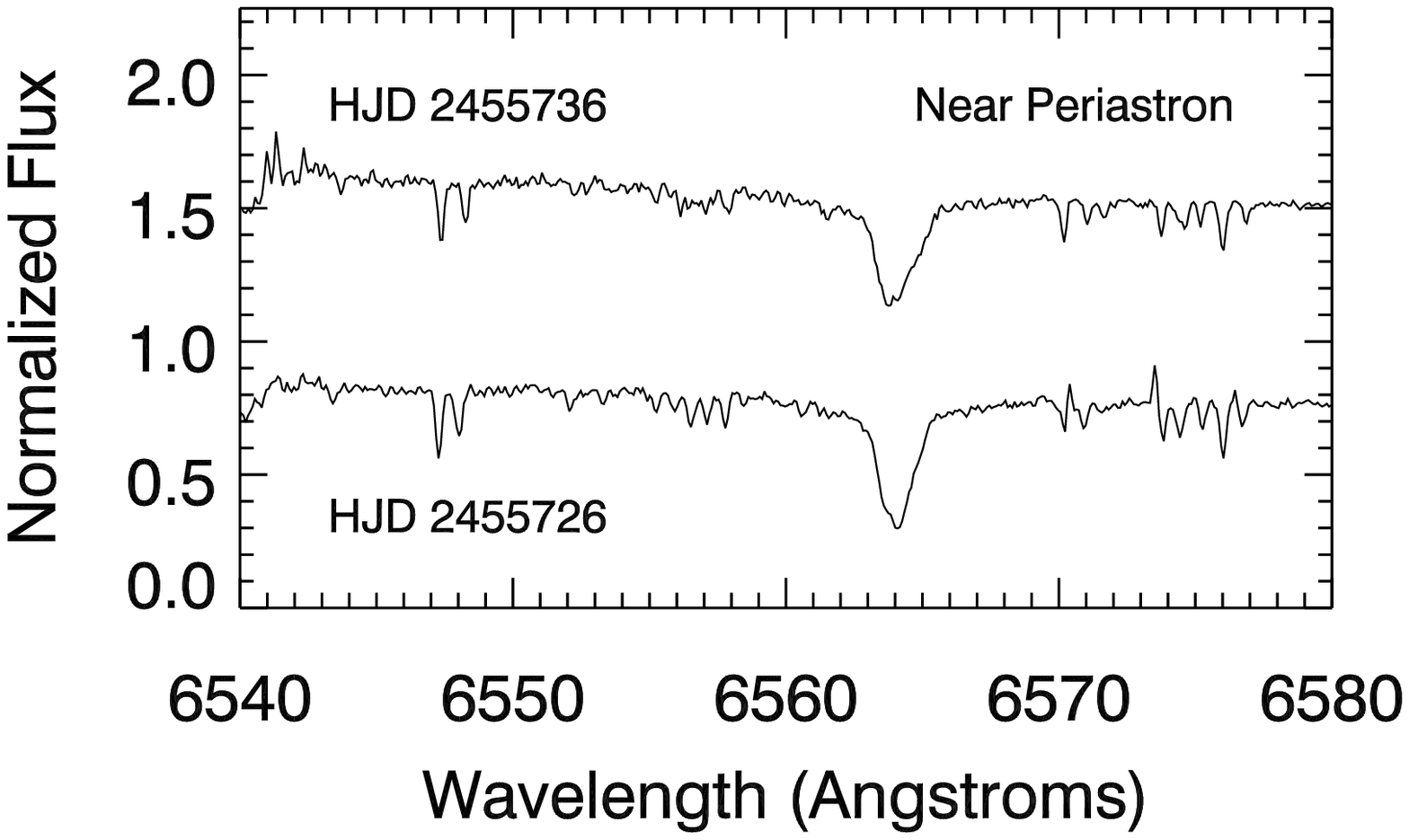}
  \caption{
  (Top) Outside periastron the combined spectrum appears convincingly single-lined.
  (Bottom) Near periastron the spectrum is resolved into its double-lined components
  (with the ARCES and HRS spectrographs, but not MARVELS). We used the double-lined
  spectrum with highest S/N when we were deriving the properties of the two
  stars via spectral characterization.
  }
  \label{fig:spectra}
\end{center}
\end{figure}

\begin{figure*}
\centering
\includegraphics[width=4.3in]{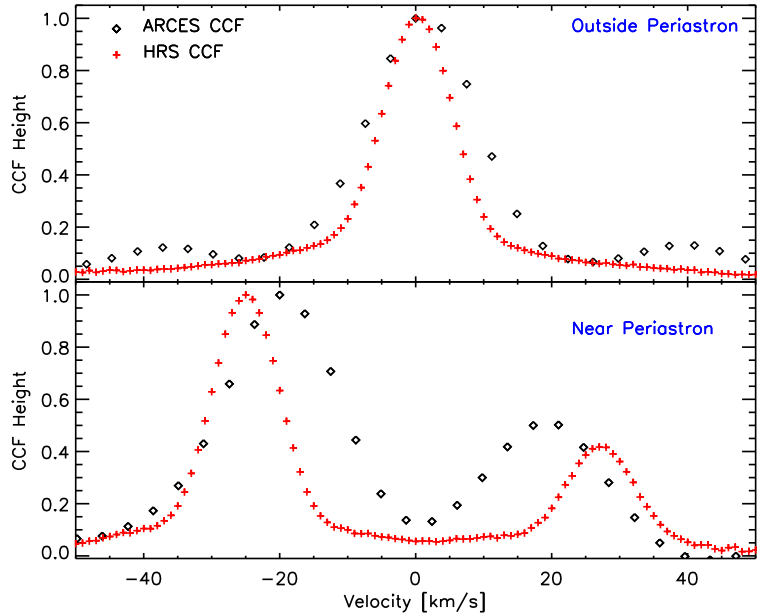}
\caption{ 
  Example CCFs obtained with the ARCES and HRS spectrographs from similar 
  (but different) phases outside of periastron (\textit{top panel}) and during periastron (\textit{bottom panel}).
  Since most of the data were obtained outside of periastron,
  most of the RV points correspond to single-peak CCFs. However, for
  data from near periastron, the ARCES and HRS spectrographs are
  able to resolve two peaks. 
  The secondary peak is comparable in height to the primary peak, 
  which led us to suspect that \tyc\ is
  an eccentric spectroscopic binary with the semi-major axis aligned perpendicular
  to the line of sight (see Figure~\ref{fig:mc10_schematic}). With this configuration, we would only resolve
  two peaks in the CCF if we happen to catch the pair of stars as they
  briefly pass through periastron. To confirm this interpretation,
  we fully observed periastron with HET/HRS, which allowed us to 
  completely constrain the orbit (see Figure~\ref{fig:mc10rv_curve_sb}).\\[5pt]
}
\label{fig:ccfs}
\end{figure*}

A total of 19 RV observations were taken with the APO 3.5m 
telescope using the ARC Echelle Spectrograph~\citep[ARCES;][]{wh03}. This spectrograph
operates in the optical regime from $\sim$3,600--10,000\AA\ with a resolving power of $R\sim31,500$.
The first set of observations were taken from 2010~October to 2011~June. 
The second set of observations, which were undertaken with the goal of increasing phase
coverage of periastron, were obtained during 2012~January--February. As shown in 
Tables~\ref{tbl:sing_rvs}~and~\ref{tbl:doub_rvs}, there were 15 ARCES points observed outside of
periastron, and 4 points during periastron (The first two of these periastron points are where we
initially resolved both the primary and secondary spectral lines---see bottom panel of 
Figures~\ref{fig:mc10rv_curve_bd},~\ref{fig:spectra},~and~\ref{fig:ccfs}---and began to suspect that the system 
might be a double-lined spectroscopic binary).

To achieve high-accuracy RV measurements with the echelle spectrograph,
we obtained a Thorium-Argon (ThAr) exposure after every science exposure.
In order to place \tyc\ on an absolute RV scale, we also frequently bracketed our
observations of \tyc\ with observations of the RV standard
HD 102158, which has an absolute RV of 28.122~km~s$^{-1}$~\citep{cj10,nm02}.
From the standard deviation of the 13 RV measurements
we obtained for HD 102158 (see Table~\ref{tbl:rvstd}), we were able to determine that the ARCES spectrograph
possesses an RV stability  of $\sim$0.5~km~s$^{-1}$. 

Two of the ARCES spectra 
were taken with longer exposure times in order
to achieve a high signal-to-noise ratio (S/N) for deriving the fundamental
stellar parameters (see Section~\ref{sec:starpars}). These two spectra were taken with an exposure time of 200 s
and with the default slit setting described in~\citet{wj12}.  The
data were reduced with IRAF, and after barycentric corrections and continuum
normalization, the two spectra were combined to produce a final spectrum
with an S/N of $\sim$170 per resolution element at $\sim$6500 \AA. However,
once we realized that \tyc\ was a double-lined spectroscopic binary, we re-derived
the spectroscopic parameters with a double-lined spectrum obtained near periastron,
as described in Section~\ref{sec:starpars}.

%%%%%%%%%%%%%%%%%%%%%%%%%%%%
%% FIGURE 2. SAMPLE SPECTRA
%%%%%%%%%%%%%%%%%%%%%%%%%%%%
%%%%%%%%%%%%%%%%%%%%%%%%%%%%

%%%%%%%%%%%%%%%%%%%%%%%%%%%
% FIGURE 3. SAMPLE CCFS
%%%%%%%%%%%%%%%%%%%%%%%%%%%
%%%%%%%%%%%%%%%%%%%%%%%%%%%%

\subsection{HET/HRS RV Data}
Upon realizing the eccentric binary-star nature of the object from the APO 3.5m data, observations where 
initiated with the 9.2m HET~\citep[][]{ra98} and the 
Higharcsec Resolution Spectrograph~\citep[HRS;][]{tr98} at a resolving power of 
\mbox{$R\sim 30,000$} using a 2 arcsec optical fiber.
A total of 18 observations were obtained 
to completely cover periastron, and thereby fully constrain the orbit.
The queue-scheduled observing mode of the
HET~\citep{sc07} is extremely well suited for investigating objects that require monitoring
over a long timespan, as well as targeted observations near periastron passage. 
For wavelength calibration, ThAr images were obtained
immediately before and after the science exposure to aid in calibrating any possible instrument drift. The data
were reduced  and wavelength calibrated using custom optimal extraction scripts written in IDL. RVs were
measured using two different techniques, which we describe below. 
The HET observations clearly resolve the
orbit for \tyc, and constrain the eccentricity to a value of $e\sim0.8$ (see Section~\ref{sec:sb2_soluxn}).

%%%%%%%%%%%%%%%%%%%%%%%%%%%%%%%%
% TABLE 3: RVSTD RVS
%%%%%%%%%%%%%%%%%%%%%%%%%%%%%%%%
  \begin{deluxetable}{ccc}%[h]
  \tablecaption{Observed heliocentric radial velocities for the RV standard HD102158\label{tbl:rvstd}}
  \tablewidth{0.38\textwidth}  
  \tablecolumns{3}
  \tablehead{\colhead{HJD} & \colhead{RV (\kms)} & \colhead{$\sigma_{\rm{RV}}$ (\kms)}}
  \startdata
  2455654.81577  & 28.734 & 0.052\\
  2455665.67180  & 28.050 & 0.034\\
  2455665.71479  & 28.169 & 0.043\\
  2455669.58461  & 27.736 & 0.045\\
  2455686.80825  & 27.470 & 0.054\\
  2455695.68591  & 28.030 & 0.038\\
  2455695.72570  & 28.090 & 0.034\\
  2455703.60341  & 28.244 & 0.039\\
  2455709.76151  & 27.476 & 0.030\\
  2455725.66705  & 26.692 & 0.123\\
  2455735.61237  & 28.126 & 0.109\\
  2455964.81467  & 28.093 & 0.058\\
  2455967.81369  & 27.728 & 0.050\\
  \enddata
  \end{deluxetable}
%%%%%%%%%%%%%%%%%%%%%%%%%%%%%%%%

\subsubsection{CCF Mask}
RVs were measured using a cross-correlation mask derived from National Solar Observatory 
Fourier transform spectroscopic solar data~\citep{ly93},
and a technique similar to that described by~\citet{bq96}.  The resultant CCF encodes information
from the $\sim$400--600 nm region, and we elected not to use redder wavelengths 
due to issues with telluric contamination.
Figure~\ref{fig:ccfs} shows the resulting CCF for an epoch during periastron and one outside of periastron; 
as is the case for the ARCES data, during periastron the primary and secondary peaks 
are clearly visible in the HET CCFs, 
but outside of periastron only a single peak is resolved.  
The centroid of the CCF peak is determined by fitting a Gaussian.

This technique has been used successfully 
for isolated stars to derive precise RVs by the teams using fiber-fed high resolution spectrographs
\citep[e.g., HARPS, SOPHIE, ELODIE, CORALIE;][]{pm00,bf06,bq96,qm00},
since PSF stability is an important component of deriving precise RVs with this technique. 
Any mismatch between the CCF and the simple Gaussian model is absorbed as a zero-point offset in
the derived RVs as long as the PSF is stable (resulting in a stable CCF shape). The HET/HRS spectrograph is also
fiber-fed, enabling this technique to also be applied to binary stars. This method is computationally efficient,
and also does not require that the spectra be normalized, resulting in a quick turn around in determining RVs
once the data are in hand. The RVs derived enabled us to plan and obtain observations as soon as the peaks began
to separate on the approach to peri-passage. Table~\ref{tbl:sing_rvs} shows the HET RVs obtained with this technique
for those epochs where the CCF appears as a single peak.

\subsubsection{TODCOR}
\label{sec:todcor}
While the CCF Mask technique described above works quite well, it does not yield the best RVs possible for 
spectra with two CCF peaks since only one mask (G2 spectral type) was used in determining peak positions.
Once all the data were in hand, we were able to apply the two-dimensional cross-correlation algorithm, TODCOR~\citep{zm94}.
TODCOR can simultaneously cross-correlate two stellar templates against a blended target stellar spectrum to disentangle the
stellar RVs of the components as well as derive a flux ratio. We used TODCOR along with HRS observations of HD161237 (G5V)
and HD 198596 (K0V) as templates to measure the RVs of \tyc. The HRS spectrum was divided into different bandpasses, and each
bandpass was solved independently following~\citet{zs03} and the resulting cross-correlation surface combined with a 
maximum likelihood analysis. Further details on our implementation of the TODCOR algorithm, as well as details of our custom
HRS spectral extraction pipeline, can be found in~\citet{bm12}.  

Table~\ref{tbl:doub_rvs} shows the RVs of the primary and secondary determined using this algorithm
at those epochs where the CCF is double peaked.
We add 0.05~km~s$^{-1}$ in quadrature to the TODCOR formal errors to account for
additional noise effects like wavelength calibration, small tracking induced PSF changes, etc.
While the HET observed the target on 18 epochs, the secondary RVs are only reliably measured for
11 epochs. These are the epochs where the primary and secondary peaks are sufficiently separated to 
determine an independent RV for each. While RVs can be determined for the other 7 epochs, they are 
RVs of blended spectra, and the associated systematic error is not only larger, but also more difficult to quantify. 

Since both peaks are unambiguously detected in TODCOR at these epochs, we  
are also able to measure the secondary to primary flux ratio, $\alpha$, which we determine 
to be  $\alpha= 0.335 \pm 0.035$ by averaging the flux ratio of the templates (G5V and K0V) 
over four bandpasses spanning 4663-5863 \AA.
Finally, the mass ratio derived from these 11 epochs is $q\sim0.88$.

\subsection{FastCam Lucky Imaging}
\label{sec:lucky}

The MARVELS team obtained lucky imaging
for \tyc\ in order to detect any spatially resolvable companions. In 2011 April,
using the FastCam~\citep{or08} instrument on the 1.5m TCS telescope at Observatorio del Teide in Spain,
we obtained 47,000 frames in the $I$-band with a 70 ms exposure time for each frame. Data processing was
accomplished with a custom-made IDL pipeline. 

\begin{figure*}
\centering
\includegraphics[width=4.3in]{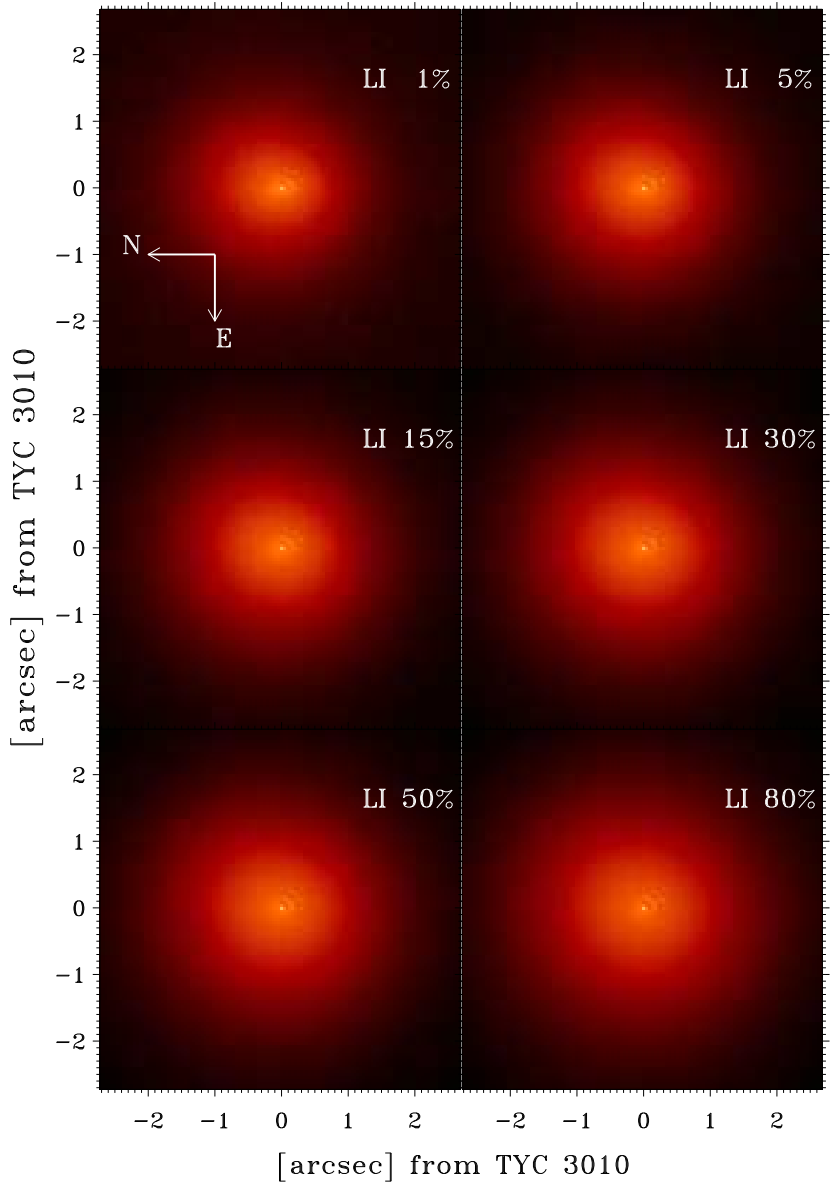}
\caption{
The best lucky imaging frames for \tyc. The best frames are selected according to the
brightest pixel (BP) method as described in Section~\ref{sec:lucky}.
}
\label{fig:lucky_exp}
\end{figure*}

As described in \citet{fg12}, the best frames are selected via the brightest pixel (BP) method.
The frames with the brightest
$X$\% of BPs are combined to generate a final image, where $X=$ \{$1,5,15,30,50,80$\} for \tyc.
Figure~\ref{fig:lucky_exp} shows the resulting final images for each particular percentage of the best frames.

No companions are detected, but we can place constraints on the upper limit of the masses of resolvable companions.
Using the spectroscopic $T_{\rm{eff}}$ for \tyc\ (see Section~\ref{sec:starpars}), 
and the relations from~\citet{ma11}, we determine the
bolometric magnitude. Combining the bolometric magnitude with mass--luminosity 
relations~\citep{hf99,ht04,df00,xr08,xf10}, we 
convert the detection
limit for the $I$-band magnitude into a lower limit for the masses of detectable companions at different
separations. At the $5\sigma$ level, where $\sigma$ is defined in~\citet{fr11}
as the rms of the counts within concentric annuli centered on \tyc, and using 8 pixel boxes,
we can rule out the presence of detectable companions above a mass of $\sim0.35\, M_{\odot}$ outside of 50~AU
(see Figure~\ref{fig:ao_mass_limit}). 

\subsection{Keck AO Imaging}
\begin{figure}[h]
\centering
\includegraphics[width=0.38\textwidth]{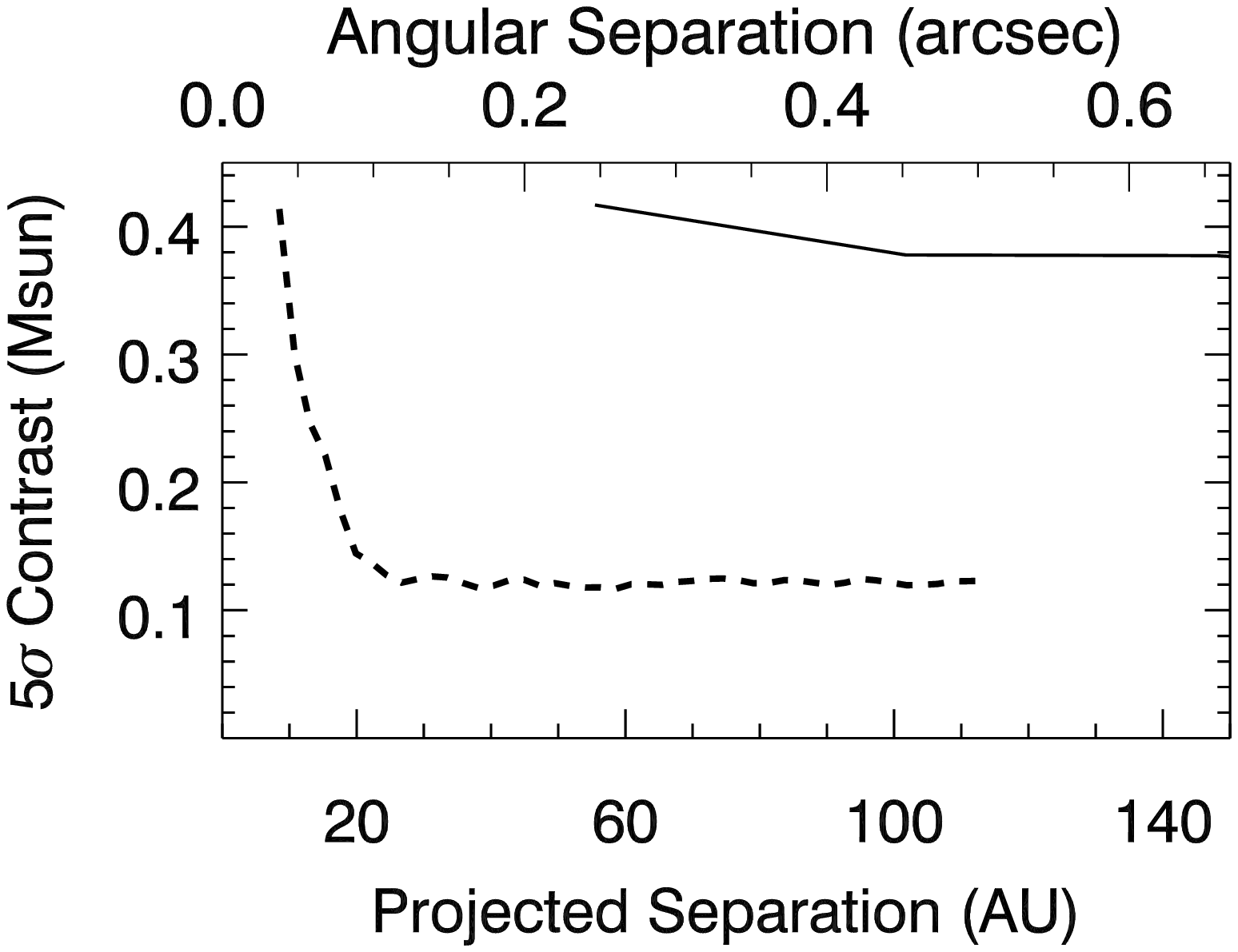}
\caption{Detectability (contrast curve) for the lucky imaging (solid) and Keck AO (dashed) images obtained
   for \tyc. Given the lucky imaging and AO detection limits, we can derive an upper limit ($5\sigma$) on the mass
   of companions as a function of angular separation.  With this upper limit, we can
   rule out the presence of companions above a mass of $\sim0.35\, M_{\odot}$ outside of $\sim$50 AU, and
   above a mass of $\sim0.13\, M_{\odot}$ outside of $\sim$100 AU.
}
\label{fig:ao_mass_limit}
\end{figure}

In addition to the lucky imaging, we were also
able to obtain adaptive optics (AO) images of \tyc\ on 2012 October 21 UT
using the NIRC2 imager at Keck~\citep[instrument PI: Keith Matthews;][]{ks94}.
Observations consist of a sequence of nine dithered frames in the $K'$
filter (central $\lambda=2.12 \mu$m) using the narrow camera (plate
scale = 10 mas pix$^{-1}$) setting. Each frame consisted of 20 coadds
with 0.1814 s of integration time per coadd, totaling 32.65 s
of on-source exposure time. Images were processed using
standard techniques to remove hot pixels, subtract the sky-background,
and align and coadd the cleaned frames. No candidate companions were
identified in either raw or processed images. Figure~\ref{fig:ao_mass_limit} shows our
sensitivity to off-axis sources as a function of angular separation.
Our diffraction-limited observations rule out the presence of
companions 6.5 magnitudes fainter than the primary star for
separations beyond 0.5$''$($5\sigma$).
Using theoretical isochrones from~\citep{gb02}, we convert this magnitude limit
to a mass upper limit, as shown in Figure~\ref{fig:ao_mass_limit};
we can exclude companions with a mass above \mbox{0.13 $M_{\odot}$} outside of 100 AU.
%%%%%%%%%%%%%%%%%%%%%%%%%%%%%%%%
%% END OF DATA
%%%%%%%%%%%%%%%%%%%%%%%%%%%%%%%%

%%%%%%%%%%%%%%%%%%%%%%%%%%%%%%%%
%% RESULTS
%%%%%%%%%%%%%%%%%%%%%%%%%%%%%%%%
\section{Results} 
\label{sec:results}
In this section we present 
the orbit solution of the \tyc\ system. First we show how the data initially suggested a spurious solution
in which \tyc\ is a single star with a BD companion. Next we present the correct solution, in which \tyc\ is shown
to be a double-lined spectroscopic stellar binary (SB2) with two solar-type stars, and we provide a full characterization
of the system properties.

%%%%%%%%%%%%%%%%%%%%%%%%%%%%
%% FIGURE 4. LUCKY IMAGES
%%%%%%%%%%%%%%%%%%%%%%%%%%%%
%%%%%%%%%%%%%%%%%%%%%%%%%%%%

\subsection{Initial spurious solution: a BD companion to a solar-type star}
\label{sec:spurious}
Of the 28 RV measurements collected with the MARVELS instrument, 24 passed the data quality
checks and were therefore included in the automated orbit solution fitting procedures. For the ARCES
data, the first 14 consecutive RV points obtained during the initial set of 
observations were fully consistent with
our working solution, that \tyc\ was a candidate BD 
(see Figure~\ref{fig:mc10rv_curve_bd} and Table~\ref{tbl:orb_params}).
These RV points are well fit by a solution consistent with a substellar object 
($M\sin{i}\sim50\, M_{\rm Jup}$) orbiting
in the BD desert around a solar-type star.
A robust fit to the low amplitude~($\sim$1--2~km~s$^{-1}$) variations 
was found with the {\sc exofast} program \citep{eg13}, which uses a set of Markov Chain Monte Carlo trials to find the best fit.
This solution, shown in Figure~\ref{fig:mc10rv_curve_bd} (top panel), is a very convincing fit to the 38 originally included
MARVELS (red points) and APO (blue points) measurements. This fit yielded a $\chi^{2}$ of 34.63 after scaling the
error bars to force $\chi^{2}$/dof$\sim$1. These scalings were not unreasonable compared to other MARVELS candidates.

As noted previously, four of the original MARVELS RV measurements were initially rejected as outliers. 
The outlier rejection procedure included a 40$\sigma$ statistical clipping to avoid phase wrapping, 
and rejection of consecutive points deviating by a large systematic offset from the bulk of the measurements. 
The latter rejection step was specifically implemented in an attempt to account for cases of fiber mis-pluggings,
which are known to happen on occasion, in which the wrong star is observed for a few observations in a row and
those few measurements appear at a very different systemic velocity relative to the majority of the measurements.
The four rejected MARVELS measurements are also shown in Figure~\ref{fig:mc10rv_curve_bd} (bottom panel, red points)
near HJD 2455250.  The final (correct) orbit solution is also shown (see details below), but it must be noted that 
this final orbit solution is only a good fit after properly disentangling the RVs from epochs
where just a single set of spectral lines is resolved; it is not a good fit to 
the {\it directly observed} single-lined RV measurements, since these are
in fact a flux-weighted average of the true primary and secondary RVs. 
The six ``outlier" measurements from this first set 
of observations (four MARVELS points and two ARCES points)
appear systematically displaced
by 15--20 \kms\ relative to the other 38 measurements, which are well fit by the spurious orbit (solid curve)
but not by the correct orbit (dashed curve).

%%%%%%%%%%%%%%%%%%%%%%%%%%%%%%
%% FIGURE 5. AO CONTRAST CURVE
%%%%%%%%%%%%%%%%%%%%%%%%%%%%%%
%%%%%%%%%%%%%%%%%%%%%%%%%%%%%%

%%%%%%%%%%%%%%%%%%%%%%%%%%%%%%%%
% TABLE 4: ORBIT PARS
%%%%%%%%%%%%%%%%%%%%%%%%%%%%%%%%
\begin{deluxetable}{lcrcr}
  \tablecaption{TYC 3010 orbital parameters: spurious\\ and true rv solutions\label{tbl:orb_params}}
  \tablecolumns{5}
  \tablewidth{0.46\textwidth}
  \tablehead{\colhead{} & \colhead{} & \colhead{Spurious solution} & \colhead{} & \colhead{True solution}}
  \startdata
                 $T_P\ (\rm{BJD_{TDB}} - 2450000)$ && $5496.8_{-2.0}^{+1.8}$ && $5970.04\pm5.1$\\
                 $P\ (\rm{days})$ && $238.49_{-0.70}^{+0.73}$ && $237.96\pm0.04$\\
                 $e$ && $0.384_{-0.048}^{+0.067}$ && $0.785\pm0.003$\\
                 $\omega\ (\rm{deg})$ && $200.88_{-2.58}^{+2.35}$ && $188.86\pm0.67$\\
                 $K_1$ \ ({\kms}) && $1.970_{-0.130}^{+0.240}$ && $15.38\pm0.25$\\
                 $K_2$ \ ({\kms}) && ... && $17.50\pm0.16$\\
                 $\gamma$ \  ({\kms}) && $61.759_{-0.087}^{+0.077}$ && $61.28\pm0.09$\\
                 $q=M_{\rm B}/M_{\rm A}$ && ... && $0.88\pm0.02$\\
  \enddata
  \tablecomments{\footnotesize
  The spurious solution consists of the {\sc exofast}~\citep{eg13} fit to the MARVELS and ARCES RV data, 
  excluding the points initially thought to be invalid outliers. The true solution was determined
  with the {\sc binary} software~\citep{gd01} and the MARVELS, ARCES, and HRS observations.
  For the true (SB2) solution, the single-lined RV measurements were disentangled
  into their primary and secondary components (see Section~\ref{sec:rv_fitting}).
  }
\end{deluxetable}
%%%%%%%%%%%%%%%%%%%%%%%%%%%%%%%%

%%%%%%%%%%%%%%%%%%%%%%%%%%%%%%%%
% TABLE 5: TYC3010 CAT PROPS
%%%%%%%%%%%%%%%%%%%%%%%%%%%%%%%%
  \begin{deluxetable}{cccc}
  \tabletypesize{\scriptsize}
  \tablewidth{0.48\textwidth}
  \tablecaption{Catalog Properties of TYC 3010-1494-1\label{tbl:mc10ab_cat_props}}
  \tablehead{\colhead{Parameter} & \colhead{Value} & \colhead{Uncertainty} & \colhead{Reference}}
  \startdata
  $\alpha$ (2000) & 11 00 11.45 &  & (1)\\
  $\delta$ (2000) & +39 43 24.74 & & (1) \\
  pmRA [mas~yr$^{-1}$] & $-43.4$ & 1.7 & (1)\\
  pmDE [mas~yr$^{-1}$] & $3.3$ & 1.6  & (1) \\
  B$_{\rm{T}}$ & $13.102$ & $0.297$ & (1) \\
  V$_{\rm{T}}$ & $11.758$ & $0.143$ & (1) \\
  $B$ & $12.007$ & $0.153$ & (2)\\
  $V$ & $11.367$ & $0.145$ & (2)\\
  $I_{\rm{C}}$ & $10.531$ & $0.074$ & (2)\\
  $g$ & $11.579$ & $0.177$ & (2)\\
  $r$ & $11.093$ & $0.089$ & (2)\\
  $i$ & $10.870$ & $0.127$ & (2)\\
  $J$ & 9.977 & 0.021 & (3)\\
  $H$ & 9.554 & 0.016 & (3)\\
  $K_{s}$ & 9.488 & 0.019 & (3)\\
  WISE1 (3.4 $\mu$m) & 9.407 & 0.006 & (4)\\
  WISE2 (4.6 $\mu$m) & 9.482 & 0.006 & (4)\\
  WISE3 (12 $\mu$m) & 9.470 & 0.038 & (4)\\
  \enddata
  \tablerefs{(1) \citet{hf00}, (2) \citet{hl12}, (3) \citet{cs03}, (4) \citet{we10}}
  \end{deluxetable}
%%%%%%%%%%%%%%%%%%%%%%%%%%%%%%%%

%%%%%%%%%%%%%%%%%%%%%%%%%%%%%%
%% FIGURE 6. SEDS
%%%%%%%%%%%%%%%%%%%%%%%%%%%%%%
  \begin{figure}[h]
  \centering
  \includegraphics[width=0.35\textwidth,angle=90]{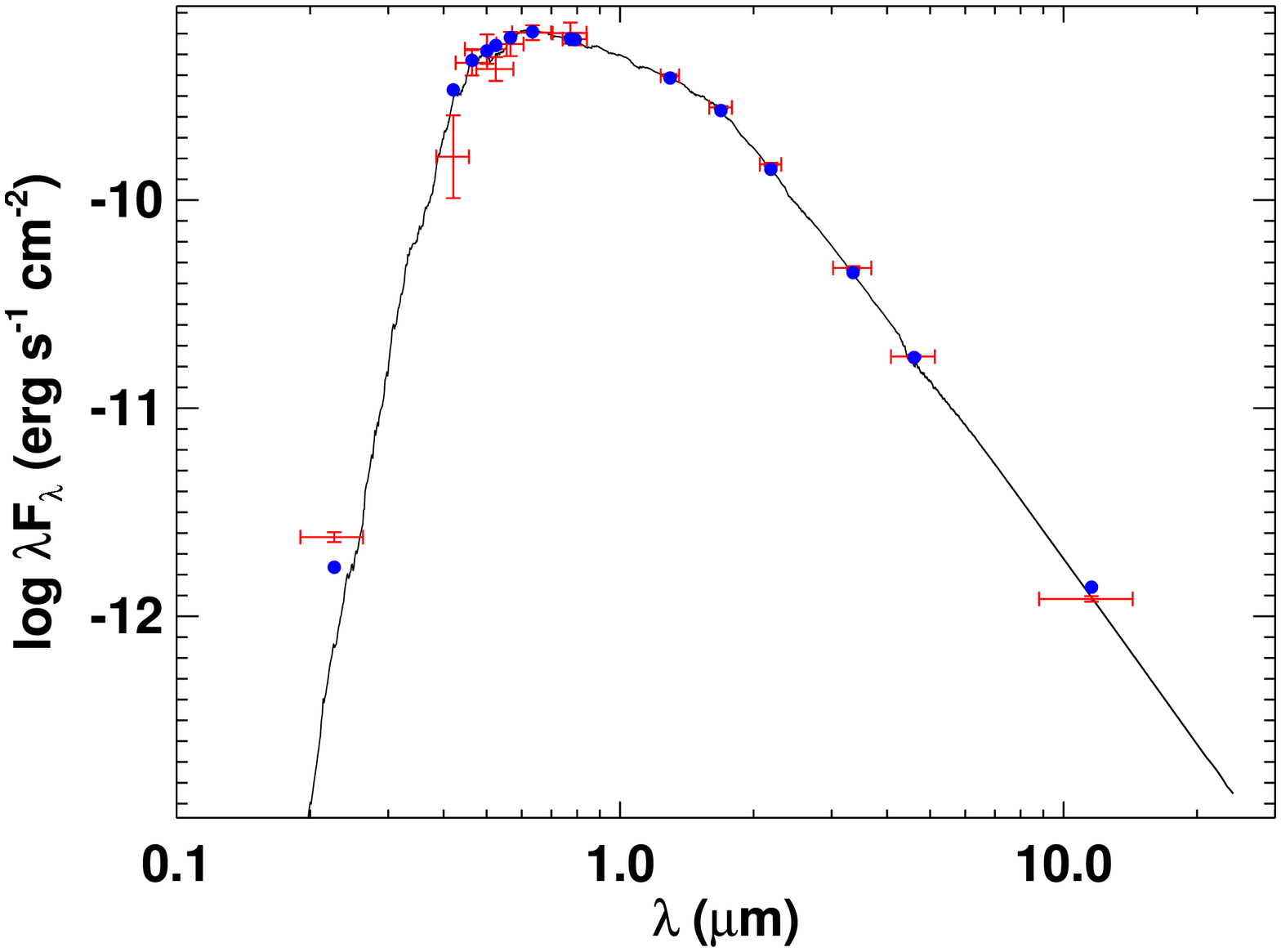}
  \includegraphics[width=0.35\textwidth,angle=90]{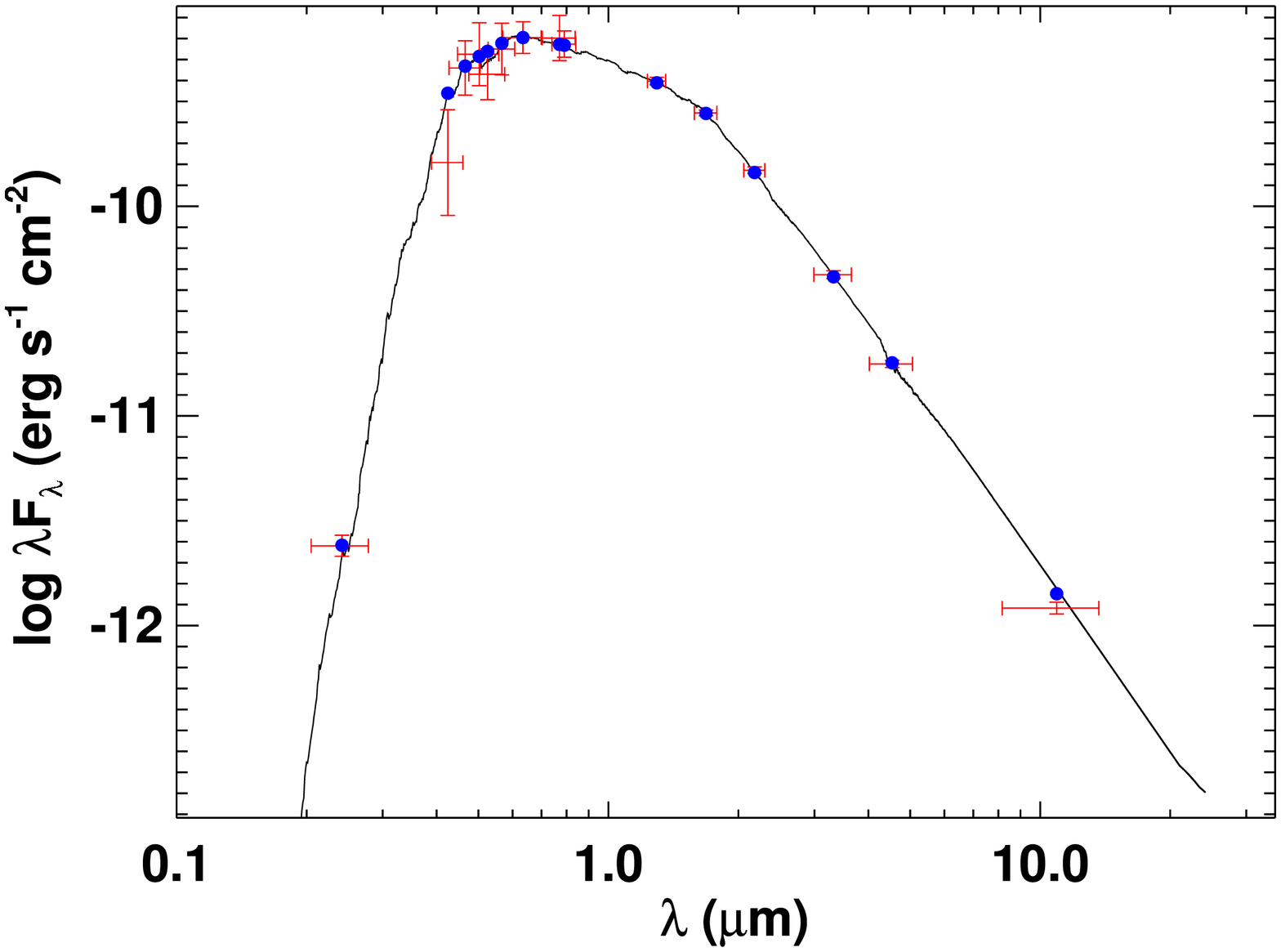}
  \caption{\textit{Top:} A NextGen model atmosphere (solid line) fit to the 
      observed broadband fluxes for TYC 3010 (assuming a single stellar component). 
      The blue points are the flux values predicted by the
      model for the different bandpasses. The vertical red bars 
      correspond to the uncertainties in the measured fluxes,
      while the horizontal red bars are the approximate widths of the bandpasses.
      This fit assumed that TYC 3010 was a single star, and found that
      $T_{\rm eff}=5400\pm100$ K, $\log{g}=4.5\pm0.5$, [Fe/H]$=0.0\pm0.1$, and $A_V=0.035\pm0.015$,
      yielding a distance of $162\pm35$ pc. 
     \textit{Bottom:} A second NextGen fit that uses two stellar components (corresponding to the primary
      and secondary stars of TYC 3010) with one of the components constrained
      to the spectroscopically determined stellar
      parameters for the primary ($T_{\rm eff}=5589\pm148$ K, $\log{g}=4.68\pm0.44$, [Fe/H]$=0.09\pm0.20$).
      This fit estimates the secondary stellar parameters to be $T_{\rm eff}=4600\pm 850$ K, 
      $R = 0.75\pm 0.4\, R_\odot$, $\log{g}= 4.6\pm 0.2$, and the distance to \tyc\
      to be $225\pm 40$ pc, with an $A_V=0.03\pm0.02$ ($\chi^2/$dof$=0.75$).\\[5pt]
      }
\label{fig:sed}
\end{figure}
%%%%%%%%%%%%%%%%%%%%%%%%%%%%%%

In addition, as we have done with all MARVELS candidates, we performed a fit to the spectral energy distribution (SED)
of the system to verify that it is consistent with a single stellar source and to provide a consistency check on the
spectroscopically determined stellar properties (see below).
We constructed the SED using fluxes (see Table~\ref{tbl:mc10ab_cat_props}) from the
Tycho catalogue~\citep{hf00}, 
APASS~\citep[\textit{A}AVSO \textit{P}hotometric \textit{A}ll-\textit{S}ky \textit{S}urvey; Data
Release 6, see][]{hl12},
Two Micron All Sky Survey~\citep{cs03}, and {\it WISE}~\citep{we10}.
NextGen models (Hauschildt et al. 1999) are used to generate
theoretical SEDs by holding $T_{\rm eff}$, $\log{g}$, and [Fe/H] at the spectroscopically determined
values (see below), and the maximum extinction $A_{\rm V}$ was limited to 0.05 mag based on the 
dust maps of~\citet{sf98}.
The best fit model can be seen in the top panel of Figure~\ref{fig:sed}; 
it corresponds to an $A_{\rm V}$ of $0.035 \pm 0.015$,
and a distance of $162 \pm 35$ pc.  
This single-star SED fit to the available photometry spanning 0.2--12$\mu$m is quite good, 
with the only hint of a discrepancy being a mild excess that appears in the {\it Galaxy Evolution Explore (GALEX)} near-UV (NUV) passband, 
despite the lack of any strong emission in the observed Ca HK lines. However, this by itself was not
deemed to be a compelling reason to suspect the high quality orbit solution.

Thus, at this point in our analysis, fully 38 RV measurements from two separate 
instruments were well fit by the same orbit solution of a single, solar-type star with a $\sim50\, M_{\rm Jup}$
companion on a modestly eccentric orbit. The SED of \tyc\ was furthermore consistent with being a single
solar-type star, and the lack of any companions in the high-resolution imaging ruled out a blend scenario 
in which the RV variations might be caused by a binary beyond 0.5$''$ of the line of sight. 
Only four of the discovery RV measurements appeared to be discrepant, and these were rejected for what appeared 
to be good reasons, behaving not unlike fiber mis-pluggings that the MARVELS team had observed in other stars before.
However, the last two RV measurements from the first set of ARCES observations appeared as strong outliers
(see Figure~\ref{fig:mc10rv_curve_bd}, blue points near HJD 2455730). 
As they were observed with a standard echelle spectrograph, these 
could not be attributed to fiber mis-pluggings, and inspection of the CCFs 
revealed double lines (see bottom panel of Figure~\ref{fig:ccfs}),
immediately nullifying the BD companion hypothesis.

%%%%%%%%%%%%%%%%%%%%%%%%%%%%%%
% FIGURE 7. SB2 RV CURVE
%%%%%%%%%%%%%%%%%%%%%%%%%%%%%%
\begin{figure}[h]
\centering
\includegraphics[width=0.45\textwidth]{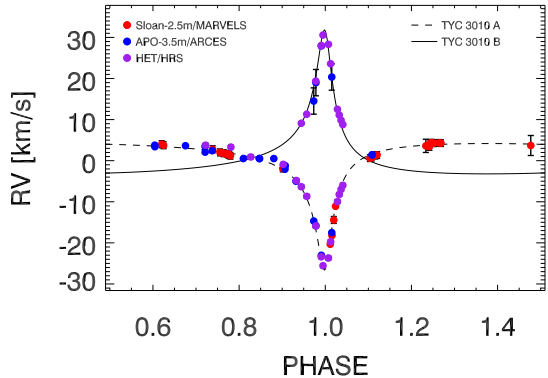}
\caption{
The correct phase-folded radial velocity curve for TYC 3010. The best-fit {\sc binary}~\citep{gd01}
orbital solution for the primary (dashed line) and secondary (solid line) are
shown with the RVs obtained from the MARVELS (red), ARCES (blue), 
and HRS (purple) spectrographs. This solution corresponds
to a period of $\sim$238 days, an eccentricity of~$\sim$0.79, with
$K_1\sim$15.38~\kms\ and $K_2\sim$17.50~\kms.  Finally,
for the RV points outside of periastron, it was necessary to de-blend the observed
RVs with the method described in Section~\ref{sec:rv_fitting}.\\[5pt]
}
\label{fig:mc10rv_curve_sb}
\end{figure}
%%%%%%%%%%%%%%%%%%%%%%%%%%%%%%

\subsection{Final solution: A highly eccentric, double-lined spectroscopic binary}
\label{sec:sb2_soluxn}
\begin{figure*}
\centering
\includegraphics[width=4.3in]{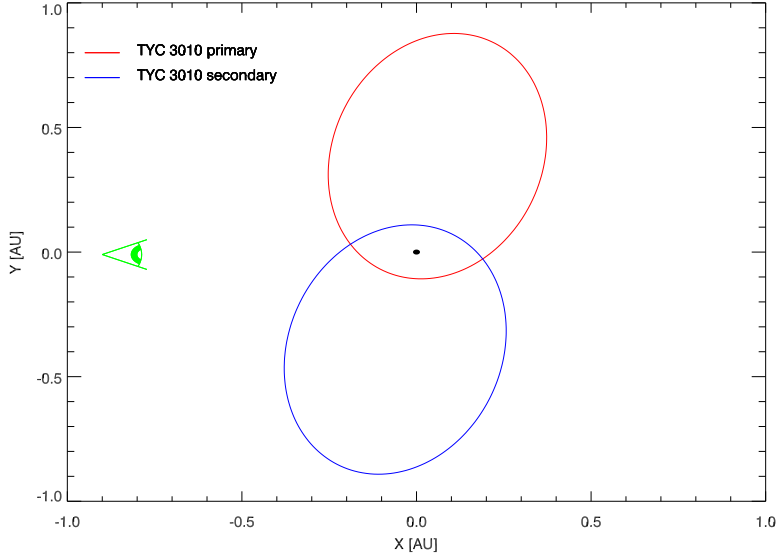}
\caption{
   A schematic of the TYC 3010 system drawn to scale, showing the primary (red) and secondary (blue)
   orbits in the orbital plane. The position of the center of mass of the system is marked by the
   black point. Given the eccentricity ($e\sim0.79$) and the fact that the semi-major
   axis is aligned nearly perpendicular to the line of sight ($\omega\sim189^{\circ}$),
   for a substantial fraction of the orbit the system can mimic
   the RV signal that would normally be induced by a secondary object with a minimum mass in the
   brown dwarf regime. Coupled with the relatively long period ($\sim$238 days), depending on
   the frequency of the observations, it can be fairly easy to miss peripassage during a
   given orbit.\\[5pt]
}
\label{fig:mc10_schematic}
\end{figure*}

To further confirm that \tyc\ was indeed a stellar binary, we closely observed the next peripassage with 
the HRS spectrograph on HET. With HET, we obtained complete coverage of periastron,
permitting a complete double-lined orbit solution. In this section we present
the correct orbit solution for \tyc, including all the points from the discovery 
and subsequent data, which shows that \tyc\ is an
SB2 with a period of $P\sim238$ days, an eccentricity of $e\sim0.79$, and 
a mass ratio of $q\sim0.88$. 
With this eccentricity and orbital period, \tyc\ lies near the upper bound of
(but within) the distribution of orbital eccentricities of solar-type binaries
with orbital periods of 100--300 days~\citep[see, e.g.,][]{dm91,rm10}.
The orbital parameters for the binary are summarized in Table~\ref{tbl:orb_params},
the RV solution is shown in Figure~\ref{fig:mc10rv_curve_sb}, and a schematic of the orbit is
shown in Figure~\ref{fig:mc10_schematic}.
In this section we also describe our determination of the stellar parameters for the primary in \tyc, 
and we estimate its mass and radius using the relations described in~\citet{ta10}. 
Since the secondary is comparable in mass to the primary, we had to take special care in accounting
for the flux contamination from the secondary, both in our determination of
the stellar parameters and with the RV values that we measured for the system outside of periastron.

\subsubsection{RV fitting}
\label{sec:rv_fitting}
For the orbital solution of the binary, we used the
RV fitting software described in~\citet{gd01}.
Since we do not resolve two sets of spectral lines for
the phases outside of periastron, most of the RV
points correspond to a flux-weighted average of the primary
and secondary RVs. In order to de-blend the flux-weighted RVs
that we measured, and derive the corresponding primary RVs,
we used the following prescription.

We treat the blended velocities as a flux-weighted average of
the primary and secondary velocities:
\begin{equation}
\label{eqn:vblend}
v_{\rm blend} = \frac{v_A F_A + v_B F_B}{F_A+F_B},
\end{equation}
where $v_A$ and $v_B$ are the primary and secondary
velocities respectively, and $F_A$ and $F_B$ are the
primary and secondary fluxes. We normalize
the flux weights by setting the sum of the fluxes, $F_A + F_B$,
to unity. Using the flux ratio, $\alpha=F_B/F_A$, from the TODCOR
analysis (which was only performed for the HET/HRS epochs where it was
possible to resolve two sets of spectral lines), we can solve for $F_A$ and $F_B$ in terms of
$\alpha$:
\begin{align}
 F_A = \frac{1}{1+\alpha}; \quad  F_B = {\alpha}{F_A}.
\end{align}
In addition, we can use the mass ratio, $q=M_B/M_A$, from the RV solution
to write $v_B$ in terms of $v_A$, since $M_B/M_A = v_A/v_B$.
\begin{align}
 v_B ={-} v_A \Big(\frac{M_A}{M_B}\Big)= \frac{-v_A}{q}
\end{align}
Returning to (\ref{eqn:vblend}), we can now write
\begin{align}
\label{eqn:deblend}
v_A = \frac{v_{\rm blend}}{F_A-F_B/q} =\Big(\frac{1+\alpha}{1-\alpha/q}\Big)\,\, v_{\rm blend}
\end{align}
With Equation~(\ref{eqn:deblend}), we can iteratively solve for
a final set of de-blended RVs for the primary. For the first iteration,
we provide an initial guess for $q$ by performing a joint fit to the
primary RVs (blended$+$unblended) combined with the secondary RVs (unblended; only
measured during periastron). Inserting this initial
guess for $q$ into Equation~(\ref{eqn:deblend}), 
we derive an initial set of de-blended
primary RVs. Then we perform another joint fit to the primary (de-blended+unblended) and secondary
(unblended) RVs to refine our value for $q$. We repeat the
process until $q$ converges. The value we find for $q$ ($0.878\pm0.016$) from this de-blending analysis is
in excellent agreement with the value for $q$ ($\sim$0.88) that we
found from the ratio of the primary and secondary RVs that were measured for the 11
HET/HRS epochs where two peaks were resolved in the CCFs. Thus, $q$ has been determined
very precisely by the orbital solution (better than 3\%), and is more precise than the
individual quoted errors on the masses.

%%%%%%%%%%%%%%%%%%%%%%%%%%%%%%
% FIGURE 8. ORBITAL SCHEMATIC
%%%%%%%%%%%%%%%%%%%%%%%%%%%%%%
%%%%%%%%%%%%%%%%%%%%%%%%%%%%%%%%%

As a further consistency check on $\alpha$ and $q$, we also note that
according to the relationship between mass and bolometric luminosity
from~\citet{ta10}, there should be a relationship between
$\alpha$ and $q$.  Since $\alpha$ is derived from a set of finite
wavelength bands, it is not bolometric. However, since the stars have temperatures
that are not too dissimilar, $\alpha$ is likely to be approximately equal to the ratio of the
bolometric luminosities. For stars with $M=0.6-1.2\, M_{\odot}$, a fit to
the~\citet{ta10} data yields $L \propto M^{5.1}$.  Thus, $\alpha=q^{5.1}$,
so $q\sim(0.335)^{1/5.1}\sim0.81$, which is within 3$\sigma$ of the value
obtained from the RV analysis.

\subsubsection{Determining the stellar parameters for \tyc}
\label{sec:starpars}
The stellar parameters for the primary were determined with
a double-lined spectrum obtained near periastron (see Section~\ref{sec:arces_data}).
The spectroscopic analysis used to determine the atmospheric parameters is similar to the one described in
\citet{wj12}, where we use two independent methods that require the conditions of excitation and 
ionization equilibria for \ion{Fe}{1} and \ion{Fe}{2} lines. These methods are referred to as the 
``BPG'' (Brazilian Participation Group) method and
the ``IAC'' (Instituto de Astrof\'{i}sica de Canarias) method.

%%%%%%%%%%%%%%%%%%%%%%%%%%%%%%%%
% TABLE 6: TYC3010 SYS PROPS
%%%%%%%%%%%%%%%%%%%%%%%%%%%%%%%%
  \begin{deluxetable}{crr}
  \tablecaption{\tyc\ properties derived by this work\label{tbl:mc10ab_sys_props}}
  \tablewidth{0.38\textwidth}
  \tablecolumns{3}
  \tablehead{\colhead{} & \colhead{System Properties} & \colhead{}}
  \startdata
  Parameter &  Value & Uncertainty\\
  \midrule
  $\alpha=F_{\rm B}/F_{\rm A}$ & 0.335 & 0.035\\
  $q=M_{\rm B}/M_{\rm A}$ & 0.878 & 0.016\\
  A$_{V}$ & 0.03 & 0.02\\
  $d$ (pc) & 225 & 40\\
  \midrule
   & \tyc\ A & \tyc\ B \\
  \midrule
  $T_{\rm{eff}}$ (K) & $5589\pm148$ & $4600\pm850$\\
  $\log{g}$ (cgs) & $4.68\pm0.44$ & $4.60\pm0.20$\\
  $[$Fe/H$]$ & $0.09\pm0.20$ & ... \\
  $M$ ($M_{\odot}$) & $1.04_{-0.12}^{+0.15}$ & $0.73_{-0.23}^{+0.24}$\\
  $R$ ($R_{\odot}$) & $0.75_{-0.27}^{+0.54}$ & $0.68_{-0.18}^{+0.23}$
  \enddata
  \tablecomments{
   The properties for the primary were determined by the spectroscopic stellar
   parameters and the~\citet{ta10} relations. The properties for the secondary were determined from the
   stellar parameters found by the two-component fit to the SED and the Torres relations.
   }
  \end{deluxetable}
%%%%%%%%%%%%%%%%%%%%%%%

The ``BPG'' analysis was done in local thermodynamic equilibrium (LTE)
using the 2002 version of {\sc moog}\footnote{http://www.as.utexas.edu/$\sim$chris/moog.html}~\citep{sc73}
and one-dimensional plane-parallel model atmospheres interpolated from the {\sc odfnew} grid of \mbox{{\sc atlas9}}
models~\citep{kr93,ck04}. In previous MARVELS papers~\citep[e.g.,][and references therein]{wj12},
the equivalent widths (EWs) of the Fe lines
were determined in an automated fashion. However, in this case, the EWs were manually measured to
carefully account for visible blends on the Fe lines from the secondary's spectrum. We note that contaminations
from very weak lines could have affected the EW measurements.
In order to correct the EWs measured for the primary
for the veiling from the continuum flux of the secondary star, we followed a
procedure similar to the one described in Section 5.2.1 of~\citet{gb08}. 
According to their prescription, we can relate the value of the true equivalent width (EW$_{\rm true}$) of a given line to the
observed equivalent width (EW$_{\rm obs}$) through the following relationship,
\begin{equation}
{\rm EW_{true, A}}= f_{\rm A}\, ({\rm EW_{obs, A}})
\end{equation}
where $f_{\rm A}$ is the so-called veiling factor for the primary.
The veiling factors for the two components are related by
\begin{equation}\label{eqn:veiling}
 \frac{f_{\rm B}(\lambda)}{f_{\rm A}(\lambda)} = \frac{F_{\rm A}(\lambda)}{F_{\rm B}(\lambda)} = \frac{1}{\alpha},
\end{equation}
where $F_{\rm A}$ and $F_{\rm B}$ are the fluxes for the primary and secondary.
Furthermore, the veiling factors satisfy the equation
\begin{equation}\label{eqn:veiling_norm}
 \frac{1}{f_{\rm A}(\lambda)} + \frac{1}{f_{\rm B}(\lambda)} = 1
\end{equation}
To simplify our analysis, we treated the veiling factors
and flux ratio as if they were wavelength independent. 
Using the average flux ratio derived by 
TODCOR ($\alpha=F_{B}/F_{A}=0.335\pm0.035$; see Section~\ref{sec:todcor}),
and the added constraint from Equation~\ref{eqn:veiling_norm}, 
we find the veiling factor for the 
primary to be $f_{\rm A}\sim1.34$. Thus, after correcting
the EWs,
we find the stellar parameters to be
$T_{\rm eff}=5589\pm148$ K, $\log{g}=4.68\pm0.44$, and [Fe/H]$=0.09\pm0.20$ (see
Table~\ref{tbl:mc10ab_sys_props}). The uncertainties for these parameters
are larger than the typical errors that we achieve with our spectroscopic analysis
because of the flux contamination from the secondary star.

The ``IAC'' analysis
extracted the stellar parameters of the
primary and secondary stars  by
considering veiling factors that were wavelength-dependent.
These veiling factors are estimated using
low-resolution Kurucz fluxes~\citep[][and references therein]{all00}
and the following equation:
\begin{equation}\label{eqn:veiling_surf}
 \frac{f_{\rm B}(\lambda)}{f_{\rm A}(\lambda)} = 
\frac{\Gamma_{\rm A}(\lambda)}{\Gamma_{\rm B}(\lambda)}\Big(\frac{R_{\rm A}}{R_{\rm B}}\Big)^2,
\end{equation}
where $\Gamma_{\rm A}$ and $\Gamma_{\rm B}$ correspond to the 
surface brightness of the primary and the secondary respectively.
To determine the ratio of the radii, we derived an empirical mass--radius relationship from
a sample of 55 stars from~\citet{ta10}, with the masses restricted to
0.7 M$_{\odot} < M <$ 1.4 M$_{\odot}$. We fit a function to
the data of the form
\begin{equation}
\log{R/R_{\odot}} = a \log(M/M_{\odot}) + b,\\[5pt]
\end{equation}
where $a=1.052\pm0.097$ and $b=0.036\pm0.008$.
Thus, the ratio of the radii for the components of \tyc\ can be written as
\begin{equation}
R_{\rm A}/R_{\rm B}=\Big(M_{\rm A}/M_{\rm B}\Big)^{1.052}
\end{equation}
The mass ratio was determined from the TODCOR analysis to be $q=M_{B}/M_{A}\sim0.88$, 
so we find that $R_{A}/R_{B}=1.142$.

As a first guess, we adopt the above values to estimate
the stellar mass and radius of the primary~\citep{all04,red06,ram07},
from solar-scaled theoretical isochrones~\citep{ber94}.
The mass ratio allows us to derive a first guess of the
$T_{\rm eff,B}$ value for the secondary to be roughly 5100~K,
assuming $\log g \sim 4.70$ and the same metallicity as
the primary.
The stellar radii we get from the comparison with isochrones are
0.89~$R_\odot$ and 0.77~$R_\odot$, and thus the ratio is
$R_A/R_B=1.145$, which is very similar to the value previously
estimated ($R_A/R_B = 1.142$).
Thus, the derived veiling factors lie in the range
$f_{\lambda,A} \sim 1.45-1.55$ and $f_{\lambda,B} \sim 3.20-2.85$
in the spectral region 4500--7000~{\AA}.

We then measure automatically, using the code
{\scshape ARES}~\citep{ss07}, the EWs of the
\ion{Fe}{1} and \ion{Fe}{2} lines~\citep{ss08} for both stellar
components and correct them using the wavelength-dependent veiling factors.
We then use the code {\scshape StePar}~\citep{tab12}
to automatically derive the stellar parameters of each component
and we get $T_{{\rm eff},A} = 5410\pm124$~K, $\log g_{A} = 4.57\pm0.56$, 
[Fe/H]$_{A} = 0.02\pm0.20$ and $\xi_{A} = 0.90\pm0.22$
from 162~\ion{Fe}{1} and 18~\ion{Fe}{2} lines.
The uncertainties are unexpectedly large and may be due to the
contamination of neighboring lines of other elements of the companion
star. Thus the results for the secondary are fairly tentative and the
errors are even larger. We were only able to measure
64~\ion{Fe}{1} and 3~\ion{Fe}{2} lines to get
$T_{{\rm eff},B} = 5136\pm323$~K, $\log g_{B} = 4.71\pm0.88$, 
[Fe/H]$_{B} = -0.15\pm0.26$ and $\xi_{B} = 0.75\pm0.40$.
Compared to the ``BPG'' analysis, the lower $T_{{\rm eff},A}$ 
of the primary may be related to the different
methods used to derive the veiling factors. 
Nevertheless, the ``IAC'' stellar parameters for the primary star are very similar to
those previously derived and are actually consistent within the large
uncertainties so we decide to adopt the ``BPG'' values.

%%%%%%%%%%%%%%%%%%%%%%%%%%%%%%%%%
% FIGURE 9. TORRES MASS & RADIUS
%%%%%%%%%%%%%%%%%%%%%%%%%%%%%%%%%
\begin{figure}[h]
\centering
\includegraphics[width=0.38\textwidth]{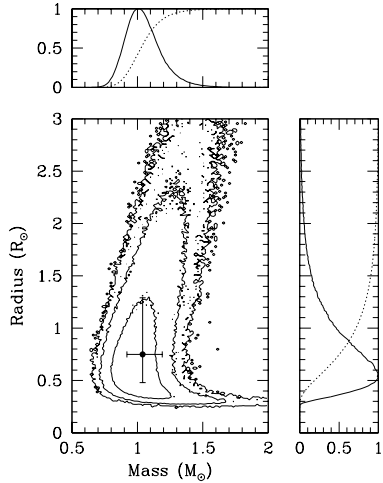}
\caption{Mass and radius distributions for the primary component of \tyc. These distributions were
determined by a set of MCMC trials with the spectroscopic stellar parameters
and the empirical relations from~\citet{ta10}. The black point represents
the median ($M_{\star}=1.04_{-0.12}^{+0.15}\, M_{\odot}$, $R_{\star}=0.75_{-0.27}^{+0.54}\, R_{\odot}$), 
and the error bars correspond to the 68.27\% confidence
intervals. The contours are lines of equal probability density
which enclose 68\%, 90\%, and 95\% of the cumulative probability
relative to the maximum of the probability density.
In the top and right panels, the probability distribution (solid line) and cumulative
probability (dashed line) are shown for the mass and radius respectively.}
\label{fig:mc10_mass_radius}
\end{figure}
%%%%%%%%%%%%%%%%%%%%%%%%%%%%%%%%%

With the ``BPG'' stellar parameters for the \tyc\ primary, we again performed a fit to the observed
SED of the system as in Section~\ref{sec:spurious}, but now also including the contribution of the secondary star.
Once again, NextGen models~\citep{ha99} are used to generate
theoretical SEDs by holding $T_{\rm eff}$, $\log{g}$, and [Fe/H] at the spectroscopically determined
values for the primary, while the $T_{\rm eff}$ for the secondary is found by the value that
minimizes $\chi^2$ ($\chi^2/$dof$=0.75$). The best fit model
can be seen in the bottom panel of Figure~\ref{fig:sed}; it corresponds 
to an $A_{\rm V}$ of $0.03\pm 0.02$,
and a distance of $225 \pm 40$ pc.
Compared to the SED fit performed in Section~\ref{sec:spurious}, which 
assumed a single stellar contribution, this two-component
SED fit no longer exhibits an excess in the {\it GALEX} NUV passband, and more generally is an excellent fit
to all of the available photometry. Finally, 
from this two-component fit to the SED, we also obtain a set of values for the
stellar parameters of the secondary of TYC 3010. We find that
$T_{\rm eff}=4600\pm850$ K, $\log{g}=4.6\pm0.2$, and [Fe/H]$=0.05\pm0.19$.

%%%%%%%%%%%%%%%%%%%%%%%%%%%%%%%%%
%% FIGURE 10. HR DIAGRAM
%%%%%%%%%%%%%%%%%%%%%%%%%%%%%%%%%
\begin{figure}[h]
\centering
\includegraphics[width=0.34\textwidth,angle=90]{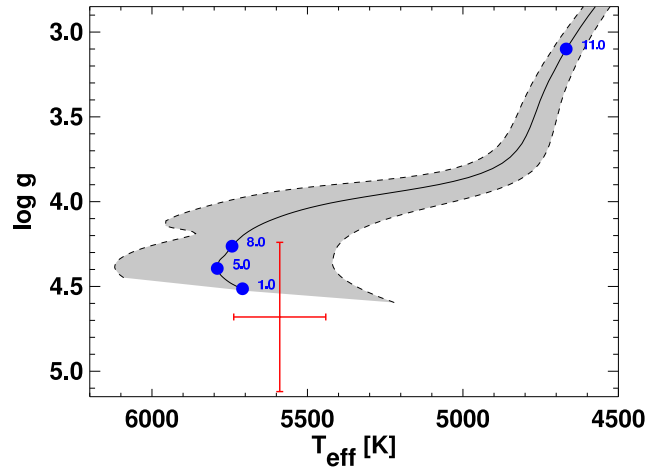}
\caption{ 
H--R diagram that compares the derived stellar parameters for the primary of \tyc\ (red error bars) to a Yonsei-Yale
stellar evolutionary track~\citep[solid curve;][]{dw04} for a star with a mass of 1.04~$M_{\odot}$ and
[Fe/H]$=0.09$. Ages (in Gyr) of 1.0, 5.0, 8.0, and 11.0 are represented by blue dots, and the 1$\sigma$ deviations
from the evolutionary track are shown in the shaded region.%\protect\\[5pt]
}
\label{fig:y2_models}
\end{figure}
%%%%%%%%%%%%%%%%%%%%%%%%%%%%%%%%%

\subsection{Inferred evolutionary status of \tyc}
Given the spectroscopic stellar parameters, we can derive the mass and
radius of the \tyc\ primary star using the empirical relationships described
in~\citet{ta10}. Figure~\ref{fig:mc10_mass_radius} shows the result of a set
of MCMC trials
for the best estimate of the mass and radius.
For the precise parameters of the primary
($T_{\rm eff}=5589$ K, $\log{g}=4.68$, [Fe/H]=0.09),
the Torres relations give 0.98~$M_{\odot}$ and 0.75~$R_{\odot}$.
Once one includes the fairly large uncertainties in the stellar parameters,
the median values for the mass and radius become $1.04_{-0.12}^{+0.15}\, M_{\odot}$ and 
$0.75_{-0.27}^{+0.54}\, R_{\odot}$, respectively. The means are $1.05\pm0.15\, M_{\odot}$ and $0.90\pm0.54\, R_{\odot}$, 
so the distributions are quite skewed as shown in Figure~\ref{fig:mc10_mass_radius}.
Compared to a Yonsei-Yale evolutionary track (see Figure~\ref{fig:y2_models}), 
we do not have a strong constraint on the
age, but \tyc\ is unlikely to have evolved off the main sequence.

We can also derive the mass and radius for the secondary given the stellar
parameters determined from the two-component SED fit and the~\citet{ta10} relations.
We find that $M_{B}=0.74_{-0.23}^{+0.26}\, M_{\odot}$ and $R_{B}=0.76_{-0.19}^{+0.27}\, R_{\odot}$.
This value for the mass of the secondary agrees within 1$\sigma$ of the value that
can be derived using the primary mass we determined above and the mass
ratio from the RV solution, i.e., $M_{B}\sim0.89\, M_{\odot}$.  

%%%%%%%%%%%%%%%%%%%%%%%%%%%%%%%%
%% END OF RESULTS
%%%%%%%%%%%%%%%%%%%%%%%%%%%%%%%%

%%%%%%%%%%%%%%%%%%%%%%%%%%%%%%%%
%% DISCUSSION
%%%%%%%%%%%%%%%%%%%%%%%%%%%%%%%%
\section{Discussion}
\label{sec:discussion}

\subsection{Why we initially derived a spurious solution}
The RV signal from \tyc\ initially seemed to indicate that it was a BD orbiting a solar-type star
in the BD desert. Over 80\% of the MARVELS discovery data agreed with this interpretation, and there seemed
to be plausible reasons for excluding the outliers. However, once similar outliers were found in the subsequent
observations, we began to suspect the validity of the BD interpretation. In this section, 
we discuss in detail why we initially favored the BD interpretation, as well as how this conclusion
was abruptly overturned by a few surprising data points.

In the discovery data, there were four outliers in total, each offset by $\sim$20 \kms\ from the
rest of the data. 
The most anomalous of the outliers was extracted from a spectrum with a 
low S/N, 
so its RV value did not seem trustworthy. 
The remaining outliers
(considering that they corresponded to a $\sim$20 \kms\ offset in RV that was
only captured once during the three orbits contained in the discovery data),
also seemed likely to be spurious.  
The MARVELS spectrograph is a fiber-fed spectrograph that can observe 60 objects simultaneously.
Each fiber is plugged by hand to observe the correct target, and occasionally a mistake may occur. 
Indeed, the MARVELS data vetting procedures were evolved to specifically include an outlier
rejection step that sought to mitigate such errors, by searching for consecutive strings of measurements
that were offset from the bulk of the data in a similar fashion to how these four measurements behave.

Remarkably, excluding these few apparent ``outliers''---and in fact {\it only} by excluding them---permits
a convincing orbit solution. It is not intuitive that this should be the case, in particular because only
$\sim$15\% of the measurements are excluded (including both the discovery data and the initial follow-up data which
appeared to corroborate the spurious solution) and because the resulting solution is so dramatically different
from the true solution. Evidently, a system such as \tyc\ (with its extreme eccentricity, leading to punctuated large
RV excursions, and its orbital orientation being nearly perpendicular to the line of sight, leading to very small
RV variations for $\sim$95\% of the orbit) is able to mimic a more circular orbit of a low-mass companion
about a single star. Moreover, the similarity of the two stars in \tyc\ leads to a combined light SED that is only
slightly different from that of a single star at a nearer distance. 

%%%%%%%%%%%%%%%%%%%%%%%%%%%%%%%%%
%% FIGURE 11. WATCH CCF FHWMS!
%%%%%%%%%%%%%%%%%%%%%%%%%%%%%%%%%
\begin{figure*}
\centering
\includegraphics[width=4.3in]{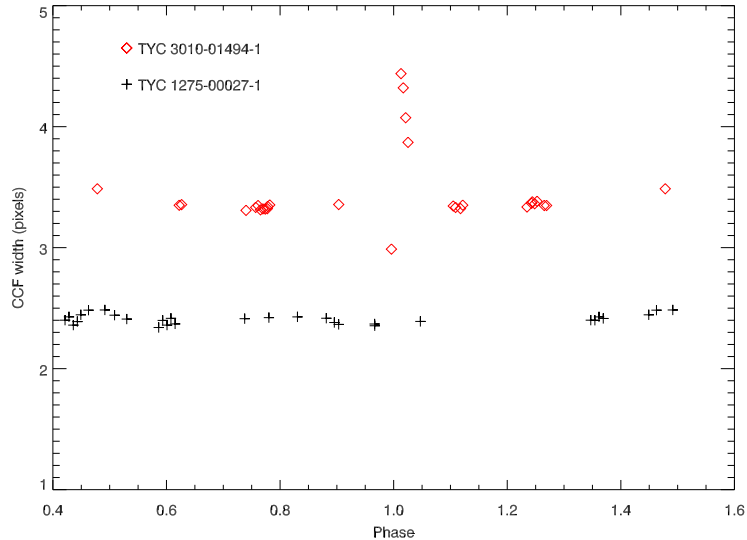}
\caption{
    A comparison of how the width of the MARVELS CCF peak varies with phase for TYC 3010 (red)
    and another MARVELS candidate, TYC 1275-00027-1 (black). The MARVELS spectrograph does
    not possess the resolution to resolve two separate peaks in the CCF for TYC 3010, even at periastron.
    Instead the width of the CCF broadens dramatically, and upon inspection 
    the peak appears asymmetric with a slight ``shoulder'' that suggests the presence of an unresolved
    secondary peak. This large variation in the peak width is not observed in TYC 1275-00027-1, which is known
    to be a single star. Therefore, by monitoring how the CCF peak changes with phase, and through visual
    inspection of the peaks, surveys can identify systems that are likely to
    be false positives like TYC 3010 during the candidate-vetting process.  Finally, the median
    value of the CCF peak width is larger for TYC 3010 than the comparison star, but this may be due to either
    TYC 3010 rotating faster or the presence of the secondary peak. When confronted with a system
    whose peak is consistently broader than one might expect for a typical solar-type star, further investigation
    is necessary to determine if it is merely a fast rotator or if it has a stellar companion.\protect\\[5pt]
}
\label{fig:ccf_widths}
\end{figure*}
%%%%%%%%%%%%%%%%%%%%%%%%

Thus many lines of evidence supported the initial solution,
considering that the BD interpretation appeared to be supported by
two years of discovery RV data, six months of additional RV observations, lucky imaging, and a well-constrained SED.
Indeed, when the two follow-up RV measurements observed near periastron appeared, indicating a possible problem with the
original orbit solution, we began to search for reasons to suspect the validity of these two anomalous points. 
At first, we thought the situation might be similar to
the fiber mis-pluggings believed to have occurred with the discovery data, and we considered that the ARCES outliers
were the result of pointing at the wrong star. But after investigating the data from those two nights, 
we confirmed that we had observed the correct target.
Next we learned of a recent change that had been made to the ARCES instrument: the ThAr
lamp had recently been replaced. The ThAr lamp is used to perform the wavelength calibration, and it was plausible
that the new lamp might have caused problems with the wavelength solution. 
Therefore, the ARCES outliers may have merely been the result 
of an artificial Doppler shift generated by an incorrect wavelength solution.
In the end, we were only able to accept that the BD interpretation was incorrect after we inspected the 
CCF for each of the outliers. The CCFs for the outliers both showed two 
peaks instead of one, indicating the presence of a second stellar component. Furthermore, 
the secondary peak was comparable in height to the primary peak (see bottom panel of Figure~\ref{fig:ccfs}),
which led us to suspect that \tyc\ was in fact a spectroscopic stellar binary.  

But how did most of the data that we had for \tyc\ conspire to imply that
it was a much less massive system? The period, shape, and orientation of the orbit with respect to the line of sight
(see Figure~\ref{fig:mc10_schematic}) made it such that for most of the orbit the two stars possess relatively low RVs 
with respect to each other. In particular, the difference between the magnitude of their RVs is smaller than the typical CCF width 
for our instruments, resulting in their CCF peaks being blended into one. Since the flux ratio is not too different
from unity, and the mass ratio is also close to unity, for epochs where the spectral lines are 
blended, there is a near-cancellation (or strong suppression)
of the true orbital velocities for the primary and secondary, which are nearly equal in magnitude but oppositely signed 
(see Equation~\ref{eqn:deblend}, and recall that $v_{\rm blend}$ is what we actually
measure). Thus, for $\sim$95\% of the orbit, the amplitude 
of the variations ($\sim$1--2 \kms) suggest a BD companion to a solar-type
star; furthermore, the eccentricity and the orbital period ensure that the stars spend 
a long time ($\sim$7 months) away from periastron, 
which is precisely the moment when the RVs of the components
are disparate enough for it to be fairly easy to resolve
the two sets of spectral lines, and the large RV amplitude ($\sim$15--20 \kms) is indicative of a stellar
binary with two solar-type stars. Moreover, the orientation makes it so that only a relatively small component of the
orbital velocities is directed along our line of sight. 
Finally, the cadence of the MARVELS survey 
made it unlikely to observe multiple epochs of periastron.

\subsection{How RV surveys can identify astrophysical false positives like \tyc}
\label{sec:lessons}
For any given RV survey, the lower the resolution of the spectrograph, the more vigilant one must be for these kinds of false
positives. For \tyc\ in particular, a spectrograph with a resolution of
$R\gtrsim{50,000}$ is required to resolve the spectral lines throughout
most of the orbit. 
But in general, as the resolution (and cadence of observations) decreases, the wider the range
of eccentricities, arguments of periastron, and orbital periods by which 
stellar binaries could masquerade
as substellar companions for significant fractions of their orbits.

Furthermore, longer period orbits ($P\gtrsim1$yr) should be handled
with special care, for in these cases the
phase coverage is more likely to be incomplete.
In order to survey $\sim$3,000 stars over four years, MARVELS required a cadence
that made it less likely to observe multiple epochs of periastron
for a binary with the period of \tyc. For MARVELS and similar RV surveys
for substellar companions, it can be costly to use precious resources
to examine false positives. Therefore, in this section,
we describe a method that the MARVELS team currently employs to 
identify binaries like \tyc\ during the candidate-vetting process.

For typical RV surveys today, a standard line bisector analysis can usually be
performed to assess the presence of blended double-lined binaries.
However, this was not possible for the MARVELS discovery data due
to its limited spectral resolution. Thus, following our experience with \tyc, 
MARVELS has developed an internal pipeline
for inspecting the widths of the CCF peaks for all of our candidates.
This way, we can readily monitor the CCFs for
signs that indicate that there
may be more than one stellar component
present (e.g., the large excursions in the width of the CCF peak that
occur near periastron for TYC 3010; see Figure~\ref{fig:ccf_widths}). There are two properties 
of the CCFs that we now monitor:
(1) the average width of the CCF peak compared to other stars in the survey,
and (2) any other significant changes in the shape of the CCF over time.

For a typical solar-type star that is
not rotating too rapidly (i.e.,
the kinds of stars that MARVELS targets),
one would expect the width of the CCF peak
to be $\sim$10 \kms, which is largely the
result of thermal broadening and micro-turbulence. However, when binary systems
like \tyc\ are unresolved, the widths of the CCF peak
are broader ($\sim$20 \kms), indicating that
there may be multiple stellar components contributing
to the flux from the system (see Figure~\ref{fig:ccf_widths}).  In fact, an atypically broad CCF peak
could also be the result of a single star rotating atypically fast, so a broad
peak is not in itself sufficient to identify the system as a binary. Nevertheless,
a broad peak should be taken as a sign to proceed with caution. 
Furthermore, changes in the skewness of
the CCF peak might provide an even more sensitive diagnostic for these kinds of
systems. Thus, by monitoring
changes in the CCF peak, even if one misses the small fraction
of the orbit where, depending on the resolution, the CCF peak 
either broadens dramatically or separates into distinct peaks
(or if one is suspicious
of the relatively few epochs where the system happened to be caught near periastron),  
it is possible to flag systems like \tyc, which may contain
much more mass than most of the RV data suggests. 

The case of \tyc\ is also a pertinent lesson on
how important it is to handle outliers carefully, especially
in this era of large surveys where thousands of objects
must be screened for the most favorable candidates. We possessed plausible
reasons for suspecting that the outliers in the discovery data might
be spurious (known issues with fiber mis-pluggings; low S/N;
and the outliers were only detected during one of the three orbits observed).
Moreover, and perhaps ironically, the spurious orbit solution is actually
a better fit to the discovery data (excluding the outliers) than the true orbit solution,
because of the need to disentangle the primary and secondary RV components from the (apparently)
single-lined RV measurements.
However, even when faced with such a compelling initial solution and
sensible reasons for considering the outliers to be invalid, it is imperative
to investigate further and provide evidence
that the reasons for rejecting the outliers are not only plausible but justified.

Furthermore, when the analysis is distributed among multiple team members
like it is within MARVELS, it is necessary to make sure each step of the
analysis is documented as clearly as possible. For MARVELS, the members
who perform the candidate-vetting are usually different from those
who perform the subsequent analysis for each candidate, so it is important
for each team member to be able to readily discover if any outliers
were rejected and why.
MARVELS has now modified its internal analysis tracking system 
in order to make the entire analysis process more transparent.

Finally, if we had been monitoring the widths of the CCF peaks, we could have
considered the evidence of the broad peak,
as well as the changing peak width around periastron, though in truth 
neither the changing width nor the broad peak by themselves
would have likely been sufficiently compelling to reject the initial orbit solution.
In the end, the most important part of our analysis
was to strategically focus our HET/HRS observations on periastron, the phase where
the outliers occurred and where it was easiest to resolve the spectral lines. This strategy
would have been more difficult with a conventionally scheduled telescope, but was readily
achieved with the queue-scheduled nature of the HET.

%%%%%%%%%%%%%%%%%%%%%%%%%%%%%%%%
%% SUMMARY
%%%%%%%%%%%%%%%%%%%%%%%%%%%%%%%%
\section{Summary}
\label{sec:summary}
We have demonstrated, using high resolution spectroscopy, that \tyc\ is an SB2. We have shown how, with a spectrograph
below a given resolution ($R\lesssim50,000$), the eccentricity and the orientation of the system with respect to our
line-of-sight allowed a large fraction of the RV curve to appear  
remarkably similar to the kind of signal one would expect
from a BD secondary as opposed to a stellar-mass secondary. Furthermore, as a result of the cadence of
the MARVELS survey and the orbital period of the system, we were more likely to miss periastron during a given
orbit. Thus, we were more susceptible to rejecting the periastron points we did obtain as outliers, even though
these points are where the spectral lines are most widely separated, and thereby where it is easiest to determine that the system is an SB2. 

Finally, we concluded with a word of warning to RV surveys, since for a given resolution and cadence, there 
are a range of orbital
parameters that can make a stellar-mass binary companion appear to be substellar. The lower the resolution or cadence,
the greater the number of stellar binaries that can masquerade in a fashion similar to \tyc. 
Therefore, if other surveys can
carefully monitor the widths of the CCF peaks for their targets (or monitor their line bisectors if they have high
enough resolution), and when possible, focus their resources
on observations of peripassage, then we hope that they will be able to avoid similar astrophysical false positives.

%%%%%%%%%%%%%%%%%%%%%%%%%%%%%%%%
%% Acknowledgments
%%%%%%%%%%%%%%%%%%%%%%%%%%%%%%%%
\acknowledgments
This research was partially supported by the Vanderbilt Initiative in Data-Intensive Astrophysics (VIDA) 
and NSF CAREER Grant AST 0349075 (CEM,KGS,LH,JP), NSF AAPF AST 08-02230 (JPW), NSF CAREER Grant AST 0645416 (EA), 
CNPq grant 476909/2006-6 (GFPM), FAPERJ grant APQ1/26/170.687/2004 (GFPM), 
NSF CAREER Grant AST-1056524 (BSG,JDE), and a PAPDRJ CAPES/FAPERJ Fellowship (LG).  

Based on observations with the SDSS 2.5-meter telescope. Funding for the MARVELS multi-object 
Doppler instrument was provided by the W.M. Keck Foundation and NSF grant AST-0705139. The MARVELS 
survey was partially funded by the SDSS-III consortium, NSF Grant AST-0705139, NASA with grant NNX07AP14G 
and the University of Florida.  The Center for Exoplanets and Habitable Worlds is supported by the 
Pennsylvania State University, the Eberly College of Science, and the Pennsylvania Space Grant Consortium.

Data presented herein were obtained
at the Hobby-Eberly Telescope (HET), a joint
project of the University of Texas at Austin, the Pennsylvania
State University, Stanford University, Ludwig-Maximilians-Universit\"{a}t M\"{u}nchen, 
and Georg-August-Universit\"{a}t G\"{o}ttingen. The HET is named in honor of
its principal benefactors, William P. Hobby and Robert
E. Eberly.

This research has made use of the SIMBAD database, operated at CDS, Strasbourg, France,
and the AAVSO Photometric All-Sky Survey (APASS), funded by the Robert Martin Ayers Sciences Fund.
It also made use of the IRAF software distributed by the National Optical Astronomy Observatory, 
which is operated by the Association of Universities for Research in Astronomy (AURA) 
under cooperative agreement with the National Science Foundation.
This publication makes use of data products from the Two Micron All Sky Survey, which is a joint project of 
the University of Massachusetts and the Infrared Processing and Analysis Center/California Institute of Technology, 
funded by the National Aeronautics and Space Administration and the National Science Foundation.
We also make use of data products from the Wide-field Infrared Survey Explorer, which is a joint project of the 
University of California, Los Angeles, and the Jet Propulsion Laboratory/California Institute of Technology, 
funded by the National Aeronautics and Space Administration.  Funding for SDSS-III has been provided by the 
Alfred P. Sloan Foundation, the Participating Institutions, the National Science Foundation, and the 
U.S. Department of Energy Office of Science.  The SDSS-III web site is http://www.sdss3.org/.

SDSS-III is managed by the Astrophysical Research Consortium for the Participating Institutions of the 
SDSS-III Collaboration including the University of Arizona, the Brazilian Participation Group, 
Brookhaven National Laboratory, University of Cambridge, University of Florida, the French Participation Group, 
the German Participation Group, the Instituto de Astrofisica de Canarias, the Michigan State/Notre Dame/JINA Participation Group, 
Johns Hopkins University, Lawrence Berkeley National Laboratory, Max Planck Institute for Astrophysics, 
New Mexico State University, New York University, Ohio State University, Pennsylvania State University, 
University of Portsmouth, Princeton University, the Spanish Participation Group, University of Tokyo, University of Utah, 
Vanderbilt University, University of Virginia, University of Washington, and Yale University.

{\it Facilities:} \facility{Sloan (MARVELS)}, \facility{ARC (ARCES)}, \facility{HET (HRS)},
\facility{Keck:I (NIRC2)}, \facility{Sanchez (FastCam)}

%%%%%%%%%%%%%%%%%%%%%%%%%%%%%%%%
%% REFERENCES
%%%%%%%%%%%%%%%%%%%%%%%%%%%%%%%%


\begin{thebibliography}{}

\bibitem[Allende Prieto et al.(2004)]{all04} Allende Prieto, C., Barklem, P.~S., Lambert, D.~L., \& Cunha, K.\ 2004, \aap, 420, 183

\bibitem[Allende Prieto \& Lambert(2000)]{all00} Allende Prieto, C., \& Lambert, D.~L.\ 2000, \aj, 119, 2445

\bibitem[{{Baranne} {et~al.}(1996){Baranne}, {Queloz}, {Mayor}, {Adrianzyk},
  {Knispel}, {Kohler}, {Lacroix}, {Meunier}, {Rimbaud}, \& {Vin}}]{bq96}
{Baranne}, A., {Queloz}, D., {Mayor}, M., {et~al.} 1996, \aaps, 119, 373

\bibitem[{{Bender} {et~al.}(2012){Bender}, {Mahadevan}, {Deshpande}, {Wright},
  {Roy}, {Terrien}, {Sigurdsson}, {Ramsey}, {Schneider}, \& {Fleming}}]{bm12}
{Bender}, C.~F., {Mahadevan}, S., {Deshpande}, R., {et~al.} 2012, \apjl, 751,
  L31

\bibitem[Bertelli et al.(1994)]{ber94} Bertelli, G., Bressan, A., Chiosi, C., Fagotto, F., \& Nasi, E.\ 1994, \aaps, 106, 275

\bibitem[{{Bouchy}(2006)}]{bf06}
{Bouchy}, F.~e. 2006, in Tenth Anniversary of 51 Peg-b: Status of and prospects
  for hot Jupiter studies, ed. L.~{Arnold}, F.~{Bouchy}, \& C.~{Moutou},
  319--325

\bibitem[{{Castelli} \& {Kurucz}(2004)}]{ck04}
{Castelli}, F., \& {Kurucz}, R.~L. 2004, ArXiv Astrophysics e-prints

\bibitem[{{Crifo} {et~al.}(2010){Crifo}, {Jasniewicz}, {Soubiran}, {Katz},
  {Siebert}, {Veltz}, \& {Udry}}]{cj10}
{Crifo}, F., {Jasniewicz}, G., {Soubiran}, C., {et~al.} 2010, \aap, 524, A10

\bibitem[{{Cutri} {et~al.}(2003){Cutri}, {Skrutskie}, {van Dyk}, {Beichman},
  {Carpenter}, {Chester}, {Cambresy}, {Evans}, {Fowler}, {Gizis}, {Howard},
  {Huchra}, {Jarrett}, {Kopan}, {Kirkpatrick}, {Light}, {Marsh}, {McCallon},
  {Schneider}, {Stiening}, {Sykes}, {Weinberg}, {Wheaton}, {Wheelock}, \&
  {Zacarias}}]{cs03}
{Cutri}, R.~M., {Skrutskie}, M.~F., {van Dyk}, S., {et~al.} 2003, VizieR Online
  Data Catalog, 2246, 0

\bibitem[{{De Lee} {et~al.}(2013)}]{dl13}
{De Lee}, N.~et~al.\ 2013, submitted

\bibitem[{{Delfosse} {et~al.}(2000){Delfosse}, {Forveille}, {S{\'e}gransan},
  {Beuzit}, {Udry}, {Perrier}, \& {Mayor}}]{df00}
{Delfosse}, X., {Forveille}, T., {S{\'e}gransan}, D., {et~al.} 2000, \aap, 364,
  217

\bibitem[{{Demarque} {et~al.}(2004){Demarque}, {Woo}, {Kim}, \& {Yi}}]{dw04}
{Demarque}, P., {Woo}, J.-H., {Kim}, Y.-C., \& {Yi}, S.~K. 2004, \apjs, 155,
  667

\bibitem[{{Duquennoy} \& {Mayor}(1991)}]{dm91}
{Duquennoy}, A., \& {Mayor}, M. 1991, \aap, 248, 485

\bibitem[{{Eastman} {et~al.}(2013){Eastman}, {Gaudi}, \& {Agol}}]{eg13}
{Eastman}, J., {Gaudi}, B.~S., \& {Agol}, E. 2013, \pasp, 125, 83

\bibitem[{{Eisenstein} {et~al.}(2011){Eisenstein}, {Weinberg}, {Agol},
  {Aihara}, {Allende Prieto}, {Anderson}, {Arns}, {Aubourg}, {Bailey},
  {Balbinot}, \& et~al.}]{ew11}
{Eisenstein}, D.~J., {Weinberg}, D.~H., {Agol}, E., {et~al.} 2011, \aj, 142, 72

\bibitem[{{Erskine}(2003)}]{ed03}
{Erskine}, D.~J. 2003, \pasp, 115, 255

\bibitem[{{Femen{\'{\i}}a} {et~al.}(2011){Femen{\'{\i}}a}, {Rebolo},
  {P{\'e}rez-Prieto}, {Hildebrandt}, {Labadie}, {P{\'e}rez-Garrido},
  {B{\'e}jar}, {D{\'{\i}}az-S{\'a}nchez}, {Vill{\'o}}, {Oscoz}, {L{\'o}pez},
  {Rodr{\'{\i}}guez}, \& {Piqueras}}]{fr11}
{Femen{\'{\i}}a}, B., {Rebolo}, R., {P{\'e}rez-Prieto}, J.~A., {et~al.} 2011,
  \mnras, 413, 1524

\bibitem[{{Fleming} {et~al.}(2010){Fleming}, {Ge}, {Mahadevan}, {Lee},
  {Eastman}, {Siverd}, {Gaudi}, {Niedzielski}, {Sivarani}, {Stassun},
  {Wolszczan}, {Barnes}, {Gary}, {Nguyen}, {Morehead}, {Wan}, {Zhao}, {Liu},
  {Guo}, {Kane}, {van Eyken}, {De Lee}, {Crepp}, {Shelden}, {Laws},
  {Wisniewski}, {Schneider}, {Pepper}, {Snedden}, {Pan}, {Bizyaev},
  {Brewington}, {Malanushenko}, {Malanushenko}, {Oravetz}, {Simmons}, \&
  {Watters}}]{fg10}
{Fleming}, S.~W., {Ge}, J., {Mahadevan}, S., {et~al.} 2010, \apj, 718, 1186

\bibitem[{{Fleming} {et~al.}(2012){Fleming}, {Ge}, {Barnes}, {Beatty}, {Crepp},
  {De Lee}, {Esposito}, {Femenia}, {Ferreira}, {Gary}, {Gaudi}, {Ghezzi},
  {Gonz{\'a}lez Hern{\'a}ndez}, {Hebb}, {Jiang}, {Lee}, {Nelson}, {Porto de
  Mello}, {Shappee}, {Stassun}, {Thompson}, {Tofflemire}, {Wisniewski},
  {Wood-Vasey}, {Agol}, {Allende Prieto}, {Bizyaev}, {Brewington}, {Cargile},
  {Coban}, {Costello}, {da Costa}, {Good}, {Hua}, {Kane}, {Lander}, {Liu},
  {Ma}, {Mahadevan}, {Maia}, {Malanushenko}, {Malanushenko}, {Muna}, {Nguyen},
  {Oravetz}, {Paegert}, {Pan}, {Pepper}, {Rebolo}, {Roebuck}, {Santiago},
  {Schneider}, {Shelden}, {Simmons}, {Sivarani}, {Snedden}, {Vincent}, {Wan},
  {Wang}, {Weaver}, {Weaver}, \& {Zhao}}]{fg12}
{Fleming}, S.~W., {Ge}, J., {Barnes}, R., {et~al.} 2012, \aj, 144, 72

\bibitem[{{Ge}(2002)}]{gj02}
{Ge}, J. 2002, \apjl, 571, L165

\bibitem[{{Ge} \& {Eisenstein}(2009)}]{ge09}
{Ge}, J., \& {Eisenstein}, D. 2009, in ArXiv Astrophysics e-prints, Vol. 2010,
  astro2010: The Astronomy and Astrophysics Decadal Survey, 86

\bibitem[{{Ge} {et~al.}(2002){Ge}, {Erskine}, \& {Rushford}}]{ge02}
{Ge}, J., {Erskine}, D.~J., \& {Rushford}, M. 2002, \pasp, 114, 1016

\bibitem[{{Ge} {et~al.}(2006){Ge}, {van Eyken}, {Mahadevan}, {DeWitt}, {Kane},
  {Cohen}, {Vanden Heuvel}, {Fleming}, {Guo}, {Henry}, {Schneider}, {Ramsey},
  {Wittenmyer}, {Endl}, {Cochran}, {Ford}, {Mart{\'{\i}}n}, {Israelian},
  {Valenti}, \& {Montes}}]{gv06}
{Ge}, J., {van Eyken}, J., {Mahadevan}, S., {et~al.} 2006, \apj, 648, 683

\bibitem[{{Ge} {et~al.}(2008){Ge}, {Mahadevan}, {Lee}, {Wan}, {Zhao}, {van
  Eyken}, {Kane}, {Guo}, {Ford}, {Fleming}, {Crepp}, {Cohen}, {Groot},
  {Galvez}, {Liu}, {Agol}, {Gaudi}, {Ford}, {Schneider}, {Seager}, {Weinberg},
  \& {Eisenstein}}]{gm08}
{Ge}, J., {Mahadevan}, S., {Lee}, B., {et~al.} 2008, in Astronomical Society of
  the Pacific Conference Series, Vol. 398, Extreme Solar Systems, ed.
  D.~{Fischer}, F.~A. {Rasio}, S.~E. {Thorsett}, \& A.~{Wolszczan}, 449

\bibitem[{{Ge} {et~al.}(2009){Ge}, {Lee}, {de Lee}, {Wan}, {Groot}, {Zhao},
  {Varosi}, {Hanna}, {Mahadevan}, {Hearty}, {Chang}, {Liu}, {van Eyken},
  {Wang}, {Pais}, {Chen}, {Shelden}, \& {Costello}}]{gl09}
{Ge}, J., {Lee}, B., {de Lee}, N., {et~al.} 2009, in Society of Photo-Optical
  Instrumentation Engineers (SPIE) Conference Series, Vol. 7440, Society of
  Photo-Optical Instrumentation Engineers (SPIE) Conference Series

\bibitem[{Girardi} {et~al.}(2002)]{gb02} Girardi, L., Bertelli, G., Bressan, A., et al.\ 2002, \aap, 391, 195

\bibitem[{{Gonz{\'a}lez Hern{\'a}ndez} {et~al.}(2008){Gonz{\'a}lez
  Hern{\'a}ndez}, {Bonifacio}, {Ludwig}, {Caffau}, {Spite}, {Spite}, {Cayrel},
  {Molaro}, {Hill}, {Fran{\c c}ois}, {Plez}, {Beers}, {Sivarani}, {Andersen},
  {Barbuy}, {Depagne}, {Nordstr{\"o}m}, \& {Primas}}]{gb08}
{Gonz{\'a}lez Hern{\'a}ndez}, J.~I., {Bonifacio}, P., {Ludwig}, H.-G., {et~al.}
  2008, \aap, 480, 233

\bibitem[{{Gudehus}(2001)}]{gd01}
{Gudehus}, D.~H. 2001, in Bulletin of the American Astronomical Society,
  Vol.~33, American Astronomical Society Meeting Abstracts \#198, 850

\bibitem[{{Gunn} {et~al.}(2006){Gunn}, {Siegmund}, {Mannery}, {Owen}, {Hull},
  {Leger}, {Carey}, {Knapp}, {York}, {Boroski}, {Kent}, {Lupton}, {Rockosi},
  {Evans}, {Waddell}, {Anderson}, {Annis}, {Barentine}, {Bartoszek}, {Bastian},
  {Bracker}, {Brewington}, {Briegel}, {Brinkmann}, {Brown}, {Carr},
  {Czarapata}, {Drennan}, {Dombeck}, {Federwitz}, {Gillespie}, {Gonzales},
  {Hansen}, {Harvanek}, {Hayes}, {Jordan}, {Kinney}, {Klaene}, {Kleinman},
  {Kron}, {Kresinski}, {Lee}, {Limmongkol}, {Lindenmeyer}, {Long}, {Loomis},
  {McGehee}, {Mantsch}, {Neilsen}, {Neswold}, {Newman}, {Nitta}, {Peoples},
  {Pier}, {Prieto}, {Prosapio}, {Rivetta}, {Schneider}, {Snedden}, \&
  {Wang}}]{gj06}
{Gunn}, J.~E., {Siegmund}, W.~A., {Mannery}, E.~J., {et~al.} 2006, \aj, 131,
  2332

\bibitem[{{Hauschildt} {et~al.}(1999){Hauschildt}, {Allard}, \& {Baron}}]{ha99}
{Hauschildt}, P.~H., {Allard}, F., \& {Baron}, E. 1999, \apj, 512, 377

\bibitem[{{Henden} {et~al.}(2012){Henden}, {Levine}, {Terrell}, {Smith}, \&
  {Welch}}]{hl12}
{Henden}, A.~A., {Levine}, S.~E., {Terrell}, D., {Smith}, T.~C., \& {Welch}, D.
  2012, Journal of the American Association of Variable Star Observers
  (JAAVSO), 40, 430

\bibitem[{{Henry}(2004)}]{ht04}
{Henry}, T.~J. 2004, in Astronomical Society of the Pacific Conference Series,
  Vol. 318, Spectroscopically and Spatially Resolving the Components of the
  Close Binary Stars, ed. R.~W. {Hilditch}, H.~{Hensberge}, \& K.~{Pavlovski},
  159--165

\bibitem[{{Henry} {et~al.}(1999){Henry}, {Franz}, {Wasserman}, {Benedict},
  {Shelus}, {Ianna}, {Kirkpatrick}, \& {McCarthy}}]{hf99}
{Henry}, T.~J., {Franz}, O.~G., {Wasserman}, L.~H., {et~al.} 1999, \apj, 512,
  864

\bibitem[{{H{\o}g} {et~al.}(2000){H{\o}g}, {Fabricius}, {Makarov}, {Urban},
  {Corbin}, {Wycoff}, {Bastian}, {Schwekendiek}, \& {Wicenec}}]{hf00}
{H{\o}g}, E., {Fabricius}, C., {Makarov}, V.~V., {et~al.} 2000, \aap, 355, L27

\bibitem[{{Jiang} {et~al.}(2013)}]{jp13}
{Jiang}, P.~et~al.\ 2013, submitted

\bibitem[{{Kurucz}(1993)}]{kr93}
{Kurucz}, R. 1993, ATLAS9 Stellar Atmosphere Programs and 2 km/s grid.~Kurucz
  CD-ROM No.~13.~ Cambridge, Mass.: Smithsonian Astrophysical Observatory,
  1993., 13

\bibitem[{{Lee} {et~al.}(2011){Lee}, {Ge}, {Fleming}, {Stassun}, {Gaudi},
  {Barnes}, {Mahadevan}, {Eastman}, {Wright}, {Siverd}, {Gary}, {Ghezzi},
  {Laws}, {Wisniewski}, {Porto de Mello}, {Ogando}, {Maia}, {Nicolaci da
  Costa}, {Sivarani}, {Pepper}, {Nguyen}, {Hebb}, {De Lee}, {Wang}, {Wan},
  {Zhao}, {Chang}, {Groot}, {Varosi}, {Hearty}, {Hanna}, {van Eyken}, {Kane},
  {Agol}, {Bizyaev}, {Bochanski}, {Brewington}, {Chen}, {Costello}, {Dou},
  {Eisenstein}, {Fletcher}, {Ford}, {Guo}, {Holtzman}, {Jiang}, {French Leger},
  {Liu}, {Long}, {Malanushenko}, {Malanushenko}, {Malik}, {Oravetz}, {Pan},
  {Rohan}, {Schneider}, {Shelden}, {Snedden}, {Simmons}, {Weaver}, {Weinberg},
  \& {Xie}}]{lg11}
{Lee}, B.~L., {Ge}, J., {Fleming}, S.~W., {et~al.} 2011, \apj, 728, 32

\bibitem[{{Lytle}(1993)}]{ly93}
{Lytle}, D.~M. 1993, in Astronomical Society of the Pacific Conference Series,
  Vol.~52, Astronomical Data Analysis Software and Systems II, ed. R.~J.
  {Hanisch}, R.~J.~V. {Brissenden}, \& J.~{Barnes}, 18

\bibitem[{{Ma} {et~al.}(2013)}]{ma13}
{Ma}, B.~et~al.\ 2013, submitted

\bibitem[Mamajek et al.\ (2011)]{ma11} Mamajek, E. 2011, Univ.\ Rochester internal memorandum. http://www.pas.rochester.edu/\\
 $\sim$emamajek/EEM\_dwarf\_UBVIJHK\_color\_Teff.dat

\bibitem[{{Mandushev} {et~al.}(2005){Mandushev}, {Torres}, {Latham},
  {Charbonneau}, {Alonso}, {White}, {Stefanik}, {Dunham}, {Brown}, \&
  {O'Donovan}}]{mt05}
{Mandushev}, G., {Torres}, G., {Latham}, D.~W., {et~al.} 2005, \apj, 621, 1061

\bibitem[{{Marcy} \& {Butler}(2000)}]{mb00}
{Marcy}, G.~W., \& {Butler}, R.~P. 2000, \pasp, 112, 137

\bibitem[{{Nidever} {et~al.}(2002){Nidever}, {Marcy}, {Butler}, {Fischer}, \&
  {Vogt}}]{nm02}
{Nidever}, D.~L., {Marcy}, G.~W., {Butler}, R.~P., {Fischer}, D.~A., \& {Vogt},
  S.~S. 2002, \apjs, 141, 503

\bibitem[{{Oscoz} {et~al.}(2008){Oscoz}, {Rebolo}, {L{\'o}pez},
  {P{\'e}rez-Garrido}, {P{\'e}rez}, {Hildebrandt}, {Rodr{\'{\i}}guez},
  {Piqueras}, {Vill{\'o}}, {Gonz{\'a}lez}, {Barrena}, {G{\'o}mez},
  {Garc{\'{\i}}a-Hern{\'a}ndez}, {Monta{\~n}{\'e}s}, {Rosenberg}, {Cadavid},
  {Calcines}, {D{\'{\i}}az-S{\'a}nchez}, {Kohley}, {Mart{\'{\i}}n},
  {Pe{\~n}ate}, \& {S{\'a}nchez}}]{or08}
{Oscoz}, A., {Rebolo}, R., {L{\'o}pez}, R., {et~al.} 2008, in Society of
  Photo-Optical Instrumentation Engineers (SPIE) Conference Series, Vol. 7014,
  Society of Photo-Optical Instrumentation Engineers (SPIE) Conference Series

\bibitem[{{Pepe} {et~al.}(2000){Pepe}, {Mayor}, {Delabre}, {Kohler}, {Lacroix},
  {Queloz}, {Udry}, {Benz}, {Bertaux}, \& {Sivan}}]{pm00}
{Pepe}, F., {Mayor}, M., {Delabre}, B., {et~al.} 2000, in Society of
  Photo-Optical Instrumentation Engineers (SPIE) Conference Series, Vol. 4008,
  Society of Photo-Optical Instrumentation Engineers (SPIE) Conference Series,
  ed. M.~{Iye} \& A.~F. {Moorwood}, 582--592

\bibitem[{{Queloz} {et~al.}(2000){Queloz}, {Mayor}, {Weber}, {Bl{\'e}cha},
  {Burnet}, {Confino}, {Naef}, {Pepe}, {Santos}, \& {Udry}}]{qm00}
{Queloz}, D., {Mayor}, M., {Weber}, L., {et~al.} 2000, \aap, 354, 99

\bibitem[{{Raghavan} {et~al.}(2010){Raghavan}, {McAlister}, {Henry}, {Latham},
  {Marcy}, {Mason}, {Gies}, {White}, \& {ten Brummelaar}}]{rm10}
{Raghavan}, D., {McAlister}, H.~A., {Henry}, T.~J., {et~al.} 2010, \apjs, 190,
  1

\bibitem[Ram{\'{\i}}rez et al.(2007)]{ram07} Ram{\'{\i}}rez, I., Allende Prieto, C., \& Lambert, D.~L.\ 2007, \aap, 465, 271

\bibitem[{{Ramsey} {et~al.}(1998){Ramsey}, {Adams}, {Barnes}, {Booth},
  {Cornell}, {Fowler}, {Gaffney}, {Glaspey}, {Good}, {Hill}, {Kelton},
  {Krabbendam}, {Long}, {MacQueen}, {Ray}, {Ricklefs}, {Sage}, {Sebring},
  {Spiesman}, \& {Steiner}}]{ra98}
{Ramsey}, L.~W., {Adams}, M.~T., {Barnes}, T.~G., {et~al.} 1998, in Society of
  Photo-Optical Instrumentation Engineers (SPIE) Conference Series, Vol. 3352,
  Society of Photo-Optical Instrumentation Engineers (SPIE) Conference Series,
  ed. L.~M. {Stepp}, 34--42

\bibitem[Reddy et al.(2006)]{red06} Reddy, B.~E., Lambert, D.~L., \& Allende Prieto, C.\ 2006, \mnras, 367, 1329

\bibitem[{{Schlegel} {et~al.}(1998){Schlegel}, {Finkbeiner}, \& {Davis}}]{sf98}
{Schlegel}, D.~J., {Finkbeiner}, D.~P., \& {Davis}, M. 1998, \apj, 500, 525

\bibitem[{{Shetrone} {et~al.}(2007){Shetrone}, {Cornell}, {Fowler}, {Gaffney},
  {Laws}, {Mader}, {Mason}, {Odewahn}, {Roman}, {Rostopchin}, {Schneider},
  {Umbarger}, \& {Westfall}}]{sc07}
{Shetrone}, M., {Cornell}, M.~E., {Fowler}, J.~R., {et~al.} 2007, \pasp, 119,
  556

\bibitem[{{Sneden}(1973)}]{sc73}
{Sneden}, C.~A. 1973, PhD thesis, THE UNIVERSITY OF TEXAS AT AUSTIN.

\bibitem[{{Sousa} {et~al.}(2007){Sousa}, {Santos}, {Israelian}, {Mayor}, \&
  {Monteiro}}]{ss07}
{Sousa}, S.~G., {Santos}, N.~C., {Israelian}, G., {Mayor}, M., \& {Monteiro},
  M.~J.~P.~F.~G. 2007, \aap, 469, 783

\bibitem[{{Sousa} {et~al.}(2008){Sousa}, {Santos}, {Mayor}, {Udry},
  {Casagrande}, {Israelian}, {Pepe}, {Queloz}, \& {Monteiro}}]{ss08}
{Sousa}, S.~G., {Santos}, N.~C., {Mayor}, M., {et~al.} 2008, \aap, 487, 373

\bibitem[Tabernero et al.(2012)]{tab12} Tabernero, H.~M., Montes, D., \& Gonz{\'a}lez Hern{\'a}ndez, J.~I.\ 2012, \aap, 547, A13

\bibitem[{{Torres} {et~al.}(2010){Torres}, {Andersen}, \& {Gim{\'e}nez}}]{ta10}
{Torres}, G., {Andersen}, J., \& {Gim{\'e}nez}, A. 2010, \aapr, 18, 67

\bibitem[{{Tull}(1998)}]{tr98}
{Tull}, R.~G. 1998, in Society of Photo-Optical Instrumentation Engineers
  (SPIE) Conference Series, Vol. 3355, Society of Photo-Optical Instrumentation
  Engineers (SPIE) Conference Series, ed. S.~{D'Odorico}, 387--398

\bibitem[{{van Eyken} {et~al.}(2010){van Eyken}, {Ge}, \& {Mahadevan}}]{vg10}
{van Eyken}, J.~C., {Ge}, J., \& {Mahadevan}, S. 2010, \apjs, 189, 156

\bibitem[{{Wang} {et~al.}(2011){Wang}, {Ge}, {Jiang}, \& {Zhao}}]{wg11}
{Wang}, J., {Ge}, J., {Jiang}, P., \& {Zhao}, B. 2011, \apj, 738, 132

\bibitem[{{Wang} {et~al.}(2012{\natexlab{a}}){Wang}, {Ge}, {Wan}, {De Lee}, \&
  {Lee}}]{wg12a}
{Wang}, J., {Ge}, J., {Wan}, X., {De Lee}, N., \& {Lee}, B. 2012{\natexlab{a}},
  \pasp, 124, 1159

\bibitem[{{Wang} {et~al.}(2012{\natexlab{b}}){Wang}, {Ge}, {Wan}, {Lee}, \& {De
  Lee}}]{wg12b}
{Wang}, J., {Ge}, J., {Wan}, X., {Lee}, B., \& {De Lee}, N. 2012{\natexlab{b}},
  \pasp, 124, 598

\bibitem[{{Wang} {et~al.}(2003){Wang}, {Hildebrand}, {Hobbs}, {Heimsath},
  {Kelderhouse}, {Loewenstein}, {Lucero}, {Rockosi}, {Sandford}, {Sundwall},
  {Thorburn}, \& {York}}]{wh03}
{Wang}, S.-i., {Hildebrand}, R.~H., {Hobbs}, L.~M., {et~al.} 2003, in Society
  of Photo-Optical Instrumentation Engineers (SPIE) Conference Series, Vol.
  4841, Society of Photo-Optical Instrumentation Engineers (SPIE) Conference
  Series, ed. M.~{Iye} \& A.~F.~M. {Moorwood}, 1145--1156

\bibitem[Matthews \& Soifer(1994)]{ks94} Matthews, K., \& Soifer, B.~T.\ 1994, 
   Astronomy with Arrays, The Next Generation, 190, 239 

\bibitem[{{Wisniewski} {et~al.}(2012){Wisniewski}, {Ge}, {Crepp}, {De Lee},
  {Eastman}, {Esposito}, {Fleming}, {Gaudi}, {Ghezzi}, {Gonzalez Hernandez},
  {Lee}, {Stassun}, {Agol}, {Allende Prieto}, {Barnes}, {Bizyaev}, {Cargile},
  {Chang}, {Da Costa}, {Porto De Mello}, {Femen{\'{\i}}a}, {Ferreira}, {Gary},
  {Hebb}, {Holtzman}, {Liu}, {Ma}, {Mack}, {Mahadevan}, {Maia}, {Nguyen},
  {Ogando}, {Oravetz}, {Paegert}, {Pan}, {Pepper}, {Rebolo}, {Santiago},
  {Schneider}, {Shelden}, {Simmons}, {Tofflemire}, {Wan}, {Wang}, \&
  {Zhao}}]{wj12}
{Wisniewski}, J.~P., {Ge}, J., {Crepp}, J.~R., {et~al.} 2012, \aj, 143, 107

\bibitem[{{Wright} {et~al.}(2010){Wright}, {Eisenhardt}, {Mainzer}, {Ressler},
  {Cutri}, {Jarrett}, {Kirkpatrick}, {Padgett}, {McMillan}, {Skrutskie},
  {Stanford}, {Cohen}, {Walker}, {Mather}, {Leisawitz}, {Gautier}, {McLean},
  {Benford}, {Lonsdale}, {Blain}, {Mendez}, {Irace}, {Duval}, {Liu}, {Royer},
  {Heinrichsen}, {Howard}, {Shannon}, {Kendall}, {Walsh}, {Larsen}, {Cardon},
  {Schick}, {Schwalm}, {Abid}, {Fabinsky}, {Naes}, \& {Tsai}}]{we10}
{Wright}, E.~L., {Eisenhardt}, P.~R.~M., {Mainzer}, A.~K., {et~al.} 2010, \aj,
  140, 1868

\bibitem[{{Wright} {et~al.}(2013)}]{wj13}
{Wright}, J.~et~al.\ 2013, submitted 

\bibitem[{{Xia} \& {Fu}(2010)}]{xf10}
{Xia}, F., \& {Fu}, Y.-N. 2010, \caa, 34, 277

\bibitem[{{Xia} {et~al.}(2008){Xia}, {Ren}, \& {Fu}}]{xr08}
{Xia}, F., {Ren}, S., \& {Fu}, Y. 2008, \apss, 314, 51

\bibitem[{{Zucker}(2003)}]{zs03}
{Zucker}, S. 2003, \mnras, 342, 1291

\bibitem[{{Zucker} \& {Mazeh}(1994)}]{zm94}
{Zucker}, S., \& {Mazeh}, T. 1994, \apj, 420, 806

\end{thebibliography}
\end{document}